\documentclass[10pt, a4paper]{article}
\usepackage[inner=2.5cm, outer=2.5cm, bottom=2.75cm, top=2.5cm]{geometry}

\AtBeginDocument{%
  \providecommand\BibTeX{{%
    \normalfont B\kern-0.5em{\scshape i\kern-0.25em b}\kern-0.8em\TeX}}}

\usepackage{amsmath}
\usepackage{amssymb}
\usepackage{amsthm}

\usepackage{mathtools}
\usepackage{stmaryrd}
\usepackage{graphicx}
\usepackage{enumitem}
\usepackage[makeroom]{cancel}
\usepackage{centernot,booktabs}
\usepackage{gensymb}
\usepackage{marginnote, cite}

\usepackage{physics}
\usepackage{qcircuit}
\usepackage[ruled,vlined]{algorithm2e}
\usepackage{tikz}
\usetikzlibrary{scopes}

\newcounter{mynote}

\newcommand{\myfintsetp}[1]{    
    \ensuremath{\left\{0, 1, \ldots, 2^{#1}-1\right\}}
}

\usepackage{float}

\newtheorem{theorem}{Proposition}[section]
\newtheorem{lemma}{Lemma}[section]

\theoremstyle{remark}
\newtheorem{remark}{Remark}[section]

\newcommand{\myrom}[1]{\uppercase\expandafter{\romannumeral #1\relax}}

\newcommand{\symmg}[1]{\mathsf{S}_{#1}} 

\definecolor{mygray}{gray}{0.3}

\DeclareFontFamily{U}{mathx}{\hyphenchar\font45}
\DeclareFontShape{U}{mathx}{m}{n}{
      <5> <6> <7> <8> <9> <10>
      <10.95> <12> <14.4> <17.28> <20.74> <24.88>
      mathx10
      }{}
\DeclareSymbolFont{mathx}{U}{mathx}{m}{n}
\DeclareFontSubstitution{U}{mathx}{m}{n}
\DeclareMathAccent{\widecheck}{0}{mathx}{"71}

\newcommand{\myprobz}{\mathfrak{P}_{\mathbf{0}}}
\newcommand{\myoutertp}[3]{\ket{#1}^{\otimes #3}\bra{#2}^{\otimes #3}}
\newcommand{\myouter}[2]{\ket{#1}\bra{#2}}
\newcommand{\mysymmsetztwo}[1]{\mathcal{S}_{#1}(\mathbb{Z}_2)}
\newcommand{\myinnerp}[2]{\left\langle #1 \middle\rvert #2 \right\rangle}
\newcommand{\myadh}[1]{\mathrm{Ad}\left(H^{\otimes #1}\right)}
\newcommand{\myhatexp}[1]{\myadh{(2k+1)}\mathrm{cexp}\left(#1\right)}
\newcommand{\myiden}[1]{\mathbb{I}_{#1}}

\newcommand{\Be}{B_{\mathrm{e}}}
\newcommand{\lossqsgi}{\Psi_{\mathrm{Q}}^{\mathrm{SGI}}}
\newcommand{\lossqgi}{\Psi_{\mathrm{Q}}^{\mathrm{GI}}}
\newcommand{\lossc}{\Psi_{\mathrm{C}}}
\newcommand{\myfintset}[1]{[\![#1]\!]}  

\usepackage{hyperref}

\newcommand{\rev}[1]{{#1}}

\makeatletter
\def\@fnsymbol#1{\ensuremath{\ifcase#1\or a\or b\or
   \mathsection\or \mathparagraph\or \|\or **\or \dagger\dagger
   \or \ddagger\ddagger \else\@ctrerr\fi}}
    \makeatother

\title{\sf A Quantum Algorithm for the Sub-Graph Isomorphism Problem}

\author{Nicola Mariella \thanks{IBM Research - Ireland; \texttt{nicola.mariella@ibm.com}. N. Mariella performed most of the work while at Mastercard Ireland.}\\
Andrea Simonetto \thanks{
UMA, ENSTA Paris, Institut Polytechnique de Paris, 91120 Palaiseau, France; \texttt{andrea.simonetto@ensta-paris.fr}. A. Simonetto performed part of the work while at IBM Research - Ireland.}}

\begin{document}

\maketitle
\begin{abstract}
We propose a novel variational method for solving the sub-graph isomorphism problem on a gate-based quantum computer. The method relies (1) on a new representation of the adjacency matrices of the underlying graphs, which requires a number of qubits that scales logarithmically with the number of vertices of the graphs; and (2) on a new Ansatz that can efficiently probe the permutation space. Simulations are then presented to showcase the approach on graphs up to $16$ vertices, whereas, given the logarithmic scaling, the approach could be applied to realistic sub-graph isomorphism problem instances in the medium term.
\end{abstract}

\section{Introduction}

We propose a method for solving the {sub-graph isomorphism} problem~\cite{Bollobas1998Modern} on a gate-based quantum computer. The method relies on a new representation of the adjacency matrices of the underlying graphs and it requires a number of qubits that scales logarithmically with the number of vertices of the graphs.

Given two undirected graphs $\mathcal{A}$ and $\mathcal{B}$ (the former with equal or larger number of vertices than the latter), the sub-graph isomorphism problem is the problem of finding occurrences of graph $\mathcal{B}$ into graph $\mathcal{A}$. If the two graphs have the same number of vertices, then the problem is known as graph isomorphism, which is therefore a special case. Graph and sub-graph isomorphism problems are also often referred to as graph matching or graph pattern matching in the computer vision literature~\footnote{This is somewhat confusing since graph matching can also refer to a specific combinatorial problem finding the set of independent edges of a single graph, and we will not use this terminology for the remainder of the paper. }. 

The sub-graph isomorphism problem (under all its embodiments) has numerous applications when data can be represented as networks, and notably in graph databases, biochemistry,
computer vision, social network analysis, knowledge graph query, among many others~\cite{Conte2004,Lee2013,bonnici2013subgraph}.
Other important examples include finding patterns to detect cyber-attacks or credit card fraud~\cite{Youhuan2019}. 

Mathematically, the sub-graph isomorphism problem is known to be NP-complete~\cite{cook_subg} (whereas the graph isomorphism problem exhibits a polynomial complexity for certain
classes of graphs and it is not known to be NP-complete in general~\cite{gizppnp}), but many classical algorithms exist tailored to different real-word graphs and instances that can deal
with graphs with thousands to millions of vertices and edges~\cite{Cordella2004,Solnon2010,Wenfei2011,Lee2013,bonnici2013subgraph,Aflalo2942,Mccreesh2018,Nabti2018,Shixuan2020}.
A special class of classical algorithms, random walks~\cite{Gori2005}, has also given rise to continuous time and discrete time quantum walks~\cite{EMMS2009985,EMMS2009934}.

Quantum algorithms for graph and sub-graph isomorphism problems are less common and often not scalable in practice. Some early negative results leveraging the connection between graph isomorphism and the hidden subgroup problem are presented in~\cite{Hallgren2010,Moore2010}, which seem to indicate that this strategy may not be fruitful in practice. A more recent theoretical result in complexity theory~\cite{kulkarni2016quantum} establishes that the minimum number of operations on a graph with $N$ vertices for solving the sub-graph isomorphism problem is of the order of magnitude of $N$. 

In the approach put forward in~\cite{CALUDE201754} (as well as in the more recent paper\rev{s}~\rev{\cite{Benkner2020,Chatterjee2021}}), if the underlying graph $\mathcal{A}$ has $N$ vertices then one needs $O(N^2)$ qubits to represent the graph isomorphism problem as a QUBO (quadratic unconstrained binary optimization)  in a quantum computer. The authors of~\cite{CALUDE201754} conjecture that this is a hard constraint for this type of formulation, which immediately requires $\approx 10^{12}$ qubit machines to hope to match current classical results.  

A slightly better scaling is presented in~\cite{Kenneth2015} and in~\cite{Gaitan_2014}, with adiabatic quantum computing-based algorithms, which have an $O(N \lceil \log_{2} N \rceil)$ qubit requirement. The former still involves QUBOs but pre-compresses the Hamiltonian classically, while the latter involves $k$-qubit couplings with $k>2$, and this hampers its efficient transpilation in hardware.  

Finally, another procedure, involving both adiabatic quantum evolution and Grover's search is detailed in~\cite{li2019quantum}, but it is specifically tailored only for the graph isomorphism problem. There, the algorithm prepares $N$ states of $N$ qubits each, but it relies on having oracles to determine the eigenvalues of a specially constructed Hamiltonian. In addition, the authors claim that the worst-case computational complexity scales as $O(N^3)$.

\subsection{Contributions}

In the present paper, by employing a carefully constructed adjacency matrix representation encoding,
we provide a quantum algorithm to solve the sub-graph isomorphism problem, that: 
\begin{itemize}
\item[\emph{(i)}] needs a number of qubits that scales logarithmically in the number of vertices of graph $\mathcal{A}$ (say $N$), thereby requiring only $O(\log_{2} N)$ qubits to run; 
\item[\emph{(ii)}] it is based on a quantum circuit which is almost immediate to compile, and we offer a worst-case depth of $O(N^2)$. Note that current quantum solutions based on QUBO and QAOA\footnote{There are $q$ variables, if the QUBO matrix is full, one needs, e.g., controlled-Z entangling gates for each pair of variables, thus the $O(q^2 p)$ scaling.} would require a worst-case depth $O(q^2 p)$ for $q$ qubits (as said either $q = O(N^2)$ or $q = O(N \lceil \log_{2} N \rceil)$) and a QAOA depth of $p$;
\item[\emph{(iii)}] it spans the permutation ``space'' efficiently (via a novel Ansatz), requiring in practice a small number of parameters to find a solution, and certainly much smaller than the permutation space of dimension $N!$. 

\end{itemize}

We showcase our algorithm in practice in quantum simulations, featuring graphs of $16$ nodes, with a permutation space of the order of $10^8$ possible combinations. The simulations demonstrate the soundness of the approach and its scalability.  Despite the considered small-scale instances, {\bf we believe that our algorithm has the potential of quantum advantage in the medium term}. In fact, a 50-qubit quantum computer could encode graphs in the order of $10^{6}$ vertices (which is the current classical limit) with 50-qubit unitary matrices, while a 100-qubit machine could encode $10^{15}$-vertex graphs, well beyond what a classical algorithm can do. We believe that is a firm step towards solving NP-complete problems on quantum computers with a quantum advantage.

\subsection{Organization}

The paper is organized as follows. We present the necessary preliminaries in Section~\ref{sec:preliminaries}.
In Section~\ref{sec:formulation}, we introduce both the classical problem and all the required steps to build a loss\footnote{In this paper, we are using loss to indicate the objective or cost of the optimization problem to be solved.} that can be evaluated on a quantum computer
in logarithmic space. Specifically we proceed constructively by first defining a classical loss for the problem, then devising a unitary representation for adjacency
matrices. Subsequently, we show that the proposed representation supports comparison and permutation, thus we obtain the corresponding quantum loss.
Section~\ref{sec:problem} presents how the sub-graph isomorphism problem can then be solved on a quantum computer by defining a pertinent Ansatz,
and therefore our novel quantum algorithm. Section~\ref{sec:experiments} collects all the numerical experiments, and we conclude in Section~\ref{sec:conclusions}.  

\section{Preliminaries}\label{sec:preliminaries}

We will work with undirected graphs and their representations in terms of adjacency matrices. We define a \textit{(finite) undirected graph} \cite{Bollobas1998Modern} $\mathcal{A}$ as the ordered pair of disjoint sets $\mathcal{A}=(V_{\mathcal{A}}, E_{\mathcal{A}})$,
where $V_{\mathcal{A}}$ is the finite set of vertices of $\mathcal{A}$ and $E_{\mathcal{A}}$ is the set of elements of the form
$\{i, j\}$, representing the edge connecting vertex $i\in V_{\mathcal{A}}$ to $j\in V_{\mathcal{A}}$.
Given a graph $\mathcal{A}$ and a strict ordering on the set of vertices $V_{\mathcal{A}}$ (i.e. for all $i, j \in V_{\mathcal{A}}$ such that $i\ne j$,
either $i>j$ or $i<j$), we associate the corresponding \textit{adjacency matrix} $A$, as the $n\times n$ matrix having entries
\begin{equation}
A_{i, j}=\begin{cases}1, & \{i, j\} \in E_\mathcal{A}\\0, & \text{otherwise,}\end{cases}
\end{equation}
where $n=|V_{\mathcal{A}}|$.
Since $\{i, j\}=\{j, i\}$ for all $i, j \in V_{\mathcal{A}}$, it follows that $A_{i, j}=A_{j, i}$, that is $A=A^\top$,
also $A_{i, j} \in \mathbb{Z}_2=\{0, 1\}$, thus the adjacency matrix is symmetric and binary. Also, as commonly assumed $A_{i,i} = 0$ for all $i$'s. 
In the present context, we consider the set $\mathbb{Z}_2$ augmented with the canonical additive group structure.
The binary operator for the addition modulo $2$ is denoted with $+_2$,
also the unary operator $(- \, \cdot)$ is used to denote the additive inverse, 
in the case of $\mathbb{Z}_2$, we have $x=(-x)$ for all $x\in\mathbb{Z}_2$, so for $x, y\in\mathbb{Z}_2$, $x +_2 (-y)=x +_2 y$.
Consequently, when considering sets of matrices over $\mathbb{Z}_2$ we implicitly augment the matrices with the same structure.
Denote $\mysymmsetztwo{n}$, the set of $n\times n$ symmetric matrices with coefficients in $\mathbb{Z}_2$, then $A \in \mysymmsetztwo{n}$.
In the following, we will make use of the following lemma which can be proved by exhaustion.
\begin{lemma}
    \label{lemma:minus_one_pow_x_plus_y_equiv}
    For all $x,y\in \mathbb{Z}_2$,
    \begin{equation}
        1 - 2(x-y)^2=(-1)^{x+y}\,.
    \end{equation}
\end{lemma}

A graph $\mathcal{A}^{\prime}$ is said to be an \textit{induced sub-graph} \cite{Bollobas1998Modern} of a graph $\mathcal{A}$ if
$V_{\mathcal{A^{\prime}}} \subset V_{\mathcal{A}}$ and for all $i, j \in V_{\mathcal{A^{\prime}}}$,
$\{i, j\} \in E_{\mathcal{A}} \iff \{i, j\} \in E_{\mathcal{A^{\prime}}}$.
This means that the sub-graph $\mathcal{A}^{\prime}$ consists of a subset of the vertices of $\mathcal{A}$ and all the edges that join
the subset of vertices $V_{\mathcal{A}^{\prime}}$ in $\mathcal{A}$.
For the remainder, we will be referring with graph and sub-graph, the (finite) undirected graph and
the induced sub-graph, respectively.

In relation to matrix algebra, we indicate with $\mathcal{M}_{n}(\mathbb{C})$, the set of square matrices of dimension $n \times n$, with elements in $\mathbb{C}$ (the imaginary unit is denoted with the symbol $\imath$, so $\imath^2=-1$). For all such matrices $A\in\mathcal{M}_{n}(\mathbb{C})$, we denote $\overline{A}$ the complex conjugate of the matrix $A$, which is
the matrix whose entries are the complex conjugates $\overline{A_{i, j}}$. With $A^*$ we denote the Hermitian of the matrix $A$, i.e.,
the complex conjugate transpose $A^*=(\overline{A})^{\top}=\overline{(A^{\top})}$.
We denote with $\myiden{n}$ the $n\times n$ identity matrix, also in the case where the subscript is not explicit, we implicitly assume $n=2$,
meaning that $\mathbb{I}=\myiden{2}$.
Specified a finite set of square matrices $A_i$, the \textit{direct sum of matrices} is the block diagonal matrix whose construction is denoted with
\begin{equation}
    \bigoplus_{i=1}^n A_i = A_1 \oplus A_2 \oplus \cdots \oplus A_n =
    \begin{pmatrix}
        A_1 & \\
        & A_2 & \\
        & & \ddots\\
        & & & A_n
    \end{pmatrix}\,.
\end{equation}
We also highlight its relation with the outer and tensor products, indeed when $A_1, A_2$ have the same size,
$A_1 \oplus A_2=\myouter{0}{0} \otimes A_1 + \myouter{1}{1}\otimes A_2$ (where $\ket{0}, \ket{1}$ are the computational basis vectors for the Hilbert space representing
the state of a qubit, while $\ket{+}, \ket{-}$ are the Hadamard ones).

Given any invertible matrix $X$ and any matrix $Y$ such that the product $XYX^{-1}$ is defined, we express the conjugation operation \cite{rossmann2006lie} by $X$ with 
$\mathrm{Ad}(X)$. Its action is defined as
\begin{equation}
\mathrm{Ad}(X)Y = XYX^{-1}\,.
\end{equation}
If $\mathbb{E}$ is a vector space and $\mathsf{GL}(\mathbb{E})$ is the space of all invertible linear transformations of $\mathbb{E}$, then 
for $a, b \in \mathsf{GL}(\mathbb{E})$, we have
\begin{subequations}
\label{ad_homom}
\begin{align}
    \mathrm{Ad}(ab)=&\mathrm{Ad}(a)\mathrm{Ad}(b),\\
    \mathrm{Ad}\left(a^{-1}\right)=&\mathrm{Ad}(a)^{-1},
\end{align}
\end{subequations}
therefore, $\mathrm{Ad}$ is a group-homomorphism. Moreover, we will be using the following result 
\begin{align}
\label{adexp}
\mathrm{exp}\left(\mathrm{Ad}(X)Y\right) = \mathrm{Ad}(X)\left(\mathrm{exp}\,Y\right),
\end{align}
which also implies that $\mathrm{exp}\left(XYX^{-1}\right) = X\left(\mathrm{exp}\,Y\right)X^{-1}$.

Referring the bra--ket notation, we define the ket symbol $\ket{a}_k$ where $a \in \myfintsetp{k}$ for some positive integer $k$ as
\begin{equation}
\ket{a}_k = \ket{a_{k-1}} \otimes \ket{a_{k-2}} \otimes \cdots \otimes \ket{a_0}
\end{equation}
with $a_t \in \{0, 1\}$ the coefficients of the binary decomposition $a=a_{k-1} 2^{k-1} + a_{k-2} 2^{k-2} + \cdots + a_0$.
The factors of the tensor product, that is the $\ket{a_t}$, are the computational basis vectors for the Hilbert space representing
the state of a qubit.
Dually, the same convention applies to the bra symbol $\bra{a}_k$. Also, let $b\in \myfintsetp{k}$, having the same range
as $a$, with binary decomposition $b=b_{k-1} 2^{k-1} + b_{k-2} 2^{k-2} + \cdots + b_0$, then the inner product $\myinnerp{a}{b}_k$ expands to
\begin{align}
\left\langle a \middle| b \right\rangle_k 
= \myinnerp{a_{k-1}}{b_{k-1}} \cdot \myinnerp{a_{k-2}}{b_{k-2}} \cdots \myinnerp{a_{0}}{b_{0}}
= \prod_{t=0}^{k-1} \myinnerp{a_t}{b_t}
= \prod_{t=0}^{k-1} \delta_{a_t, b_t} = \delta_{a, b}
\end{align}
When there is no ambiguity on the range of the expression in the bra--ket symbol, we omit the subscript $k$, and we simply put $\ket{a}_k=\ket{a}$.

Given $i, j \in \myfintsetp{k}$ we define the following tensor product shortcut
\begin{align}
    \ket{i, j} \coloneqq& \ket{i} \otimes \ket{j}
    = \ket{i_{k-1}} \otimes \ket{i_{k-2}} \otimes \cdots \otimes \ket{i_0} \otimes \ket{j_{k-1}} \otimes \ket{j_{k-2}} \otimes \cdots \otimes \ket{j_0},
\end{align}
the same convention in terms of the outer product translates to
\begin{align}
    \myouter{i, j}{i, j}=\myouter{i}{i} \otimes \myouter{j}{j}
    = \myouter{i_{k-1}}{i_{k-1}} \otimes \cdots \otimes \myouter{i_0}{i_0}
    \otimes \myouter{j_{k-1}}{j_{k-1}} \otimes \cdots \otimes \myouter{j_0}{j_0}\,.
\end{align}

Given an unitary operator $U$ acting on $k$ qubits, we define $\mathsf{C}(U)$ the controlled unitary whose control qubit
is the most significant qubit, as
\begin{align}
    \mathsf{C}(U) \coloneqq& \myouter{0}{0}\otimes \mathbb{I}^{\otimes k} + \myouter{1}{1}\otimes U,
\end{align}
thus $\mathsf{C}(U)$ acts on $k+1$ qubits.
The underlying matrix of $\mathsf{C}(U)$ is the block diagonal matrix $\mathsf{C}(U) = \mathbb{I}^{\otimes k} \oplus U$.

In the context of this work, we note that if $U$ is a diagonal operator then so is $\mathsf{C}(U)$.
Also, given two arbitrary unitary operators $U_1$ and $U_2$ acting both on $k$ qubits, it follows that
\begin{subequations}
\begin{align}
    \label{cnot_decomp}
    \mathsf{C}(U_1)\cdot \mathsf{C}(U_2)
    =& \mathsf{C}(U_1\cdot U_2),\\
    \label{cnot_conj_decomp}
    \mathsf{C}(U_1 U_2 U_1^*)
    =& (\mathbb{I}^{\otimes k}\oplus U_1)\cdot \mathsf{C}(U_2) \cdot (\mathbb{I}^{\otimes k}\oplus U_1^*)\\
    = & (U_1 \oplus U_1)\cdot \mathsf{C}(U_2) \cdot (U_1^* \oplus U_1^*) \qquad (\textrm{since } U_1 U_1^* = \mathbb{I}^{\otimes k})\\
    =& (\mathbb{I}\otimes U_1)\cdot \mathsf{C}(U_2) \cdot (\mathbb{I}\otimes U_1^*)\,.
\end{align}
\end{subequations}
For unitary operator $U$ and $\mathsf{C}(U)$, the following lemma holds, which can be proven by direct computation. 

\begin{lemma}
    \label{lemma:ctrl_U_inner_prod_plus}
    For any unitary operator $U$ and state $\ket{\psi}$ both on $k$ qubits, such that $\myinnerp{\psi}{\psi}=1$ we have 
    \begin{subequations}
    \begin{align}
        \left(\bra{+}\otimes \bra{\psi}\right) \cdot \mathsf{C}(U) \cdot \left(\ket{+}\otimes \ket{\psi}\right)
        =& \left(\bra{+}\otimes \bra{\psi}\right)
        \cdot\left(\myouter{0}{0} \otimes \myiden{2^k} + \myouter{1}{1} \otimes U \right)
        \cdot \left(\ket{+}\otimes \ket{\psi}\right)\\
        =& \frac{1}{2} + \frac{1}{2}\bra{\psi}U\ket{\psi}\,.
    \end{align}
    \end{subequations}
\end{lemma}
In relation to matrix exponentiation, we introduce a shorthand notation for the controlled $\mathrm{exp}$, so given a skew-hermitian $2^k\times 2^k$ matrix $K$, we define
\begin{align}
    \mathrm{cexp}(K) \coloneqq& \mathsf{C}(\mathrm{exp}(K)) = \mathbb{I}^{\otimes k} \oplus \mathrm{exp}(K)\,.
\end{align}

We introduce a relation between the tensor product $\otimes$ and the direct sum of matrices $\oplus$ as per our definition of the latter.
First notice that given vector spaces $V$ and $W$, the tensor product has the commutative property in the sense that
$V\otimes W\cong W\otimes V$ (where $\cong$ means isomorphic to), indeed consider the linear map with rule $a\otimes b\mapsto b\otimes a$, then there exists its inverse, hence
$V\otimes W$ and $W \otimes V$ are isomorphic.
Let $A, C, D$ be linear operators, then
\begin{subequations}
\label{otimes_oplus_relation}
\begin{align}
    A\otimes \left(C\oplus D\right) =& A\otimes \left(\myouter{0}{0}\otimes C + \myouter{1}{1}\otimes D\right)\\
    \cong& \myouter{0}{0} \otimes A \otimes C + \myouter{1}{1}\otimes A\otimes D = \left(A\otimes C\right)\oplus\left(A\otimes D\right)\,.
\end{align}
\end{subequations}

In the present work, when implementing quantum measurements on $n$ qubits, we will always consider the projection operator $\mathcal{O}=\myoutertp{0}{0}{n}$.
Given a unitary $U$ acting on $n$ qubits, and assuming the initial state is $\ket{0}^{\otimes n}$, we measure the probability of observing the bitstring of
$n$ zeros, upon measuring the state $\ket{\psi}=U\ket{0}^{\otimes n}$.
We denote the probability of obtaining the latter outcome as
\begin{subequations}
\label{probz_def}
\begin{align}
    \myprobz(U) \coloneqq& \mathrm{tr}\left(\myouter{\psi}{\psi}\mathcal{O}\right) = \bra{\psi} \mathcal{O} \ket{\psi}\\
    =& \bra{0}^{\otimes n} U^* \myoutertp{0}{0}{n} U \ket{0}^{\otimes n} = \left| \bra{0}^{\otimes n} U \ket{0}^{\otimes n}\right|^2\,.
\end{align}
\end{subequations}

\section{Formulation for the Sub-graph Isomorphism Problem}\label{sec:formulation}

\subsection{The classical loss function}
\label{subsection:classical_loss}

With the preliminaries in place, we are now ready to formulate the sub-graph isomorphism problem as an optimization problem. Consider graphs $\mathcal{A} $ and $\mathcal{B}$, and identify their corresponding adjacency matrices
$A \in \mathcal{S}_{N_A}(\mathbb{Z}_2)$ and $B \in \mathcal{S}_{N_B}(\mathbb{Z}_2)$, with $N_A \ge N_B$. 

Furthermore define the block selection matrix $S = \begin{pmatrix}\mathbb{I}_{N_B} & \vert & \mathbb{O}_{N_B, N_A-N_B}\end{pmatrix}$, where $\mathbb{I}_{N_B}$ is the $N_B\times N_B$ identity matrix
and $\mathbb{O}_{N_B, N_A-N_B}$ is the $N_B \times (N_A-N_B)$ matrix of zeros.
The action of matrix $S$ on the matrix $A$, that is $SAS^\top$, is that of selecting the first $N_B \times N_B$ upper block of $A$.

We now define the permutation matrices as $P \in \Pi_N$, where $\Pi_N$ is the set of permutation matrices of dimension $N \times N$, that will transform $A$ to match the resulting upper block to $B$. The permutation matrices will be our decision variables. 
  
We then define a classical loss function $\ell_{\mathrm{C}} : \Pi_{N_A} \times \mathcal{S}_{N_A}(\mathbb{Z}_2) \times \mathcal{S}_{N_B}(\mathbb{Z}_2) \to [0, N_B^2]$ as,
\begin{equation}\label{eq:classical-loss-1}
    \ell_{\mathrm{C}}(P; A, B) = \left\|S P A P^\top S^{\top} - B \right\|^2_{\mathrm{F}}, 
\end{equation}
(where $\|\cdot\|_{\mathrm{F}}$ is the Frobenius norm), and the subgraph isomorphism problem as the combinatorial optimization problem
\begin{equation}\label{eq:sgi-classical}
   \min_{P \in \Pi_{N_A}} \, \ell_{\mathrm{C}}(P; A, B).
\end{equation}

As mentioned in the introduction, Problem~\eqref{eq:sgi-classical} is known to be NP-complete~\cite{cook_subg}, and it is solved with tailored-made heuristics up to millions
of variables classically. When $N_A = N_B$ and therefore the selection matrix $S$ is the identity, Problem~\eqref{eq:sgi-classical} becomes the graph isomorphism problem.

To better link the permutation $P$ of the adjacency matrix $A$ with the permutation of the underlying vertices of the graph, which is in fact what we are interested in, and to better characterize certain properties of the problem, it is important to introduce a vertex permutation mapping. 

In this context, we consider the \textit{defining representation}~\cite{Sagan2001} of the symmetric group $\mathsf{S}_N$ on $\{0, 1, \ldots, N-1\}=: [\![N-1]\!]$, which is of
degree $N$. Let $p \in \mathsf{S}_N$ be a permutation function in $\mathsf{S}_N$ such that $p: [\![N-1]\!] \to [\![N-1]\!]$, and we define the
(injective) group-homomorphism $\varphi: \mathsf{S}_N \to \Pi_N$ as
\begin{equation}\label{p-P}
\varphi(p) := \sum_{i=0}^{N-1}\ket{p(i)}\bra{i},
\end{equation}
which maps an element of the symmetric group to an $N\times N$ permutation matrix. Note that $\varphi(p)$ defines the map $\ket{i} \mapsto \ket{p(i)}$ for all
$i \in [\![N-1]\!]$. While this extra representation seems to be somewhat unnecessary, one can appreciate its power as a proof tool in the following useful lemma (see also Figure~\ref{fig.example} for a pictorial example). \begin{lemma}
    \label{lemma-p_plus}
    For any $N\times N$ permutation matrix $P$, with $N=2^k$,
    \begin{equation}
        P\ket{+}^{\otimes k}=\ket{+}^{\otimes k}
    \end{equation}
    that is $\ket{+}^{\otimes k}$ is an eigenvector corresponding to the eigenvalue 1.
    Equivalently, $\ket{+}^{\otimes k}$ is a state stabilized by $P$.
\end{lemma}
\begin{proof}
    Since $\ket{+}^{\otimes k}=\frac{1}{\sqrt{2^{k}}}\sum_{i=0}^{N-1}\ket{i}$, and $P$ is the image on $\varphi$ of some $p\in \mathsf{S}_N$,
    then
    \begin{equation}
        P\ket{+}^{\otimes k} = \left(\sum_{i=0}^{N-1}\ket{p(i)}\bra{i}\right) \frac{1}{\sqrt{2^{k}}}\sum_{i^{\prime}=0}^{N-1}\ket{i^{\prime}}
        = \frac{1}{\sqrt{2^{k}}}\sum_{i=0}^{N-1}\ket{p(i)},
    \end{equation}
    from the bijectivity of $p$, it follows that $\sum_{i=0}^{N-1}\ket{p(i)}=\sum_{i^{\prime}=0}^{N-1}\ket{i^{\prime}}$,
    hence $P\ket{+}^{\otimes k}=\ket{+}^{\otimes k}$, as required.
\end{proof}

Returning to our task, once function $p$ is chosen, say $\bar{p}$, we can write and determine the resulting permutation matrix $P_{\bar{p}} \in \Pi_N$ as $P_{\bar{p}} = \varphi(\bar{p})$. We can therefore rewrite Problem~\eqref{eq:sgi-classical} in terms of vertex permutation function $p$ as,
\begin{equation}\label{eq:sgi-classical-2}
   \min_{p \in \mathsf{S}_N, p: [\![N-1]\!] \to [\![N-1]\!]} \, \ell_{\mathrm{C}}(P_{p}; A, B).
\end{equation}
We will come back to these two formulations when dealing with quantum algorithms. 

In the following, we will describe how to encode Problem~\eqref{eq:sgi-classical} or equivalently \eqref{eq:sgi-classical-2} in a quantum circuit. We will proceed by steps.
First, we will show how to represent any adjacency matrix in a quantum circuit that uses a number of qubits that scales logarithmically with the nodes
of the graph. Then, we will use this representation to encode also the permutations, and build the loss function, maintaining the logarithmic scaling with respect to the number of vertices. And finally, we will show how to encode the decision variables (e.g., the $P$'s) into an optimizable Ansatz. 

\begin{remark}
We remark that the choice of the permutation as the product $SP$ may appear unnatural, specifically, 
the data needed to determine the action of $W=SP$ is indeed a sub-matrix of $P$.
To show this, let $W=SP$ be a \textit{partial permutation}, that is a $N_B\times N_A$ matrix having properties that
$WW^{\top}=SP(SP)^{\top}=SPP^{\top}S^{\top}=SS^{\top}=\mathbb{I}_{N_B}$ and the set of rows of $V$ is a subset of the set of vectors for the standard basis of $\mathbb{R}^{N_A}$,
with cardinality $N_B$.     
We consider the block matrix decomposition $P=\left({P^{\prime}}^{\top}\middle|{P^{\prime\prime}}^{\top}\right)^{\top}$, where $P^{\prime}$ and $P^{\prime\prime}$
have size $N_B\times N_A$ and $(N_A-N_B)\times N_A$, respectively. Then
\begin{align}
    W =& SP = \begin{pmatrix}\mathbb{I}_{N_B} & \vert & \mathbb{O}_{N_B, N_A-N_B}\end{pmatrix}\cdot 
    \begin{pmatrix}P^{\prime}\\P^{\prime\prime}\end{pmatrix} = P^{\prime}\,.
\end{align}

Finally, we note that the size of the search space for the $N_B$-permutation of $N_A$ is given by $\frac{N_A!}{(N_A-N_B)!}$, which becomes
$N_A!$ in the case $N_A=N_B$, that is the specific case of graph isomorphism problem.
\hfill $\Box$
\end{remark}
\smallskip

\begin{figure}
    \centering \vspace*{0.5cm}
    \scalebox{0.8}{
    \begin{tikzpicture}
        \begin{scope}[every node/.style={circle,thick,draw}]
            \node (3) at (0,0) {3};
            \node (0) at (0,2) {0};
            \node (2) at (2,2) {2};
            \node (1) at (2,0) {1};
        \end{scope}
        \begin{scope}[
            every node/.style={fill=white,circle},
            every edge/.style={draw=black,very thick}]
            \path [-] (0) edge (1);
            \path [-] (0) edge (3);
            \path [-] (1) edge (3);
            \path [-] (1) edge (2);
        \end{scope} 
        \draw[arrows=->,line width=1pt](2.5,1)--(4,1);
        \begin{scope}[every node/.style={circle,thick,draw}]
            \node (1) at (5,0) {3};
            \node (3) at (5,2) {0};
            \node (0) at (7,2) {2};
            \node (2) at (7,0) {1};
        \end{scope}
        \begin{scope}[
            every node/.style={fill=white,circle},
            every edge/.style={draw=black,very thick}]
            \path [-] (0) edge (1);
            \path [-] (0) edge (3);
            \path [-] (1) edge (3);
            \path [-] (1) edge (2);
        \end{scope}
        \node at (12.0,1) {$P = \left[\begin{array}{cccc} 
        0 & 0 & 1 & 0 \\ 
        0 & 0 & 0 & 1 \\
        0 & 1 & 0 & 0 \\
        1 & 0 & 0 & 0  \end{array}\right], \quad p = \left\{\begin{array}{c} p(0)=2\\ p(1)=3\\ p(2)=1\\ p(3) = 0 \end{array}\right.$};

    \end{tikzpicture}}
    \caption{Example for different permutation representations.}
    \label{fig.example}
\end{figure}
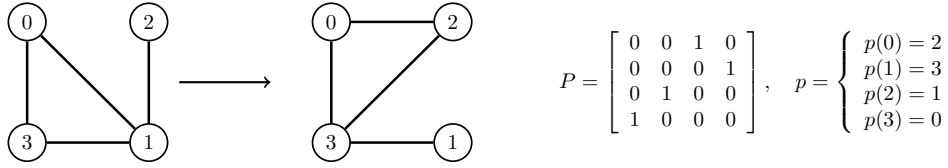

\begin{figure}
    \centering\vspace*{0.5cm}
    \scalebox{0.8}{
    \begin{tikzpicture}
        \node at (1,2.) {$\mathcal{A}$};
        \begin{scope}[every node/.style={circle,thick,draw}]
        \node (3) at (0,0) {3};
        \node (0) at (0,2) {0};
        \node (2) at (2,2) {2};
        \node (1) at (2,0) {1};
        \end{scope}
        \begin{scope}[
        every node/.style={fill=white,circle},
        every edge/.style={draw=black,very thick}]
        \path [-] (0) edge (1);
        \path [-] (0) edge (3);
        \path [-] (1) edge (3);
        \path [-] (1) edge (2);
        \end{scope} 
        \draw[thick,->] (4,1) -- (2.5,1) node[midway,above] {$f$};
        \node at (5.6,1.3) {$\mathcal{B}$};
        \begin{scope}[every node/.style={circle,thick,draw}]
        \node (1) at (5,0) {2};
        \node (2) at (5,2) {0};
        \node (0) at (7,2) {1};
        \end{scope}
        \begin{scope}[
        every node/.style={fill=white,circle},
        every edge/.style={draw=black,very thick}]
        \path [-] (0) edge (2);
        \path [-] (0) edge (1);
        \path [-] (1) edge (2);
        \end{scope}
    \end{tikzpicture}}
    \caption{The (induced) sub-graph isomorphism problem. {Here $f$ is an injective graph homomorphism that preserves the connectivity,
    in the sense that for all $\{i, j\} \in V_{\mathcal{B}}$, we have $f(V_{\mathcal{B}}) \subseteq V_{\mathcal{A}}$ and 
    $\{f(i), f(j)\} \in E_{\mathcal{A}} \iff \{i, j\} \in E_{\mathcal{B}}$.}}
\end{figure}
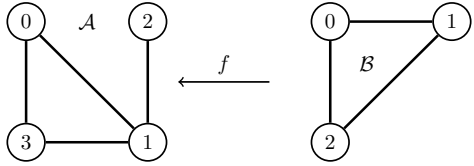

\subsection{Adjacency representation}

Consider graph $\mathcal{A}=\left(V_{\mathcal{A}}, E_{\mathcal{A}}\right)$ and its adjacency matrix $A$. Without loss of generality\footnote{
The case where the number of vertices is not a power of two can be overcome by introducing $2^{\lceil \log_2 \left|V_{\mathcal{A}}\right| \rceil} - \left|V_{\mathcal{A}}\right|$
ancillary singleton vertices. The latter corresponds to adding a number of singleton vertices such that the total number of vertices is 
$2^{\lceil \log_2 \left|V_{\mathcal{A}}\right| \rceil}$, which is the least power of two that is greater than or equal to $\left|V_{\mathcal{A}}\right|$.
}, we impose the size of the adjacency matrices to be a power of 2, that is $N\times N$, with $N=2^k$ for some non-negative integer $k$.
Then, the adjacency matrix $A$ belongs to the set $\mathcal{S}_{N}(\mathbb{Z}_2)$.

Define the injective mapping $h: \mathcal{S}_N(\mathbb{Z}_2) \to \mathcal{M}_{N^2}(\mathbb{C})$ from the set of symmetric $N\times N$ matrices with coefficients in
$\mathbb{Z}_2$ to the
set of $N^2\times N^2$ square matrices in $\mathbb{C}$ as
\begin{align}
    \label{hrule}
    A = \sum_{i, j} A_{i, j}\rvert i\rangle \langle j \rvert \: \overset{h}{\mapsto}  \: \imath\frac{2\pi}{\left| \mathbb{Z}_2 \right|} \sum_{i, j} A_{i, j} \myouter{i, j}{i, j}\,.
\end{align}

The constant $\imath \frac{2\pi}{\left| \mathbb{Z}_2 \right|}=\imath\pi$ (where we remind that $\imath$ is the imaginary unit) is the phase for the corresponding element $A_{i, j} \in \mathbb{Z}_2$,
and its purpose is that of creating a group-isomorphism between $\mathbb{Z}_2$ and the cyclic group of the square roots of
unity $\{1, -1\}$ under multiplication (as it will become clearer by the proof of Lemma~\ref{lemma:exph} below). Informally, the isomorphism reflects the `classical' addition of matrices, as composition of unitary operators
in quantum computing. The mapping $h$ can be interpreted as a flattening of a matrix in $\mathcal{S}_N(\mathbb{Z}_2)$ along the diagonal of a matrix in $\mathcal{M}_{N^2}(\mathbb{C})$,
that is the basis vectors $\rvert i\rangle \langle j \rvert$ are mapped to the diagonal basis
$\rvert i, j\rangle \langle i, j \rvert = \rvert i\rangle \langle i \rvert \otimes \rvert j\rangle \langle j \rvert$,
where the vector $\ket{i, j}$ spans $2k$ qubits. We note that $h(A)$ is skew-hermitian since $h(A)^* = - h(A)$. 
Then since $h(A)$ is diagonal for all $A \in \mathcal{S}_N(\mathbb{Z}_2)$, we see that
\begin{subequations}
\begin{align}
    \label{simple_exph_exansion}
    \mathrm{exp}\left(h(A)\right) =& \mathrm{exp}\left(\imath\pi \sum_{i, j} A_{i, j} \myouter{i, j}{i, j}\right)=\sum_{i, j} (-1)^{A_{i, j}} \myouter{i, j}{i, j},\\
    \label{simple_cexph_exansion}
    \mathrm{cexp}\left(h(A)\right) =&
    \mathbb{I}_{N^2} \oplus \sum_{i, j} (-1)^{A_{i, j}} \myouter{i, j}{i, j}
\end{align}
\end{subequations}

We establish a lemma emphasizing the action of the exponential mapping on the function $h$ introduced in \eqref{hrule}.
Its results will be key in later theoretical developments.

\begin{lemma}
    \label{lemma:exph}
    For all $A, B \in \mysymmsetztwo{N}$, the following results hold:
    \begin{subequations}
    \label{exphprops}
    \begin{align}
       \mathrm{(i)} && \mathrm{exp}\left(h(A)\right) =& \mathrm{exp}\left(h(A)\right)^*
        = \mathrm{exp}\left(h(A)^*\right)
        = \mathrm{exp}\left(\overline{h(A)}\right)
        = \mathrm{exp}\left(-h(A)\right);\label{inside1}\\
        \mathrm{(ii)} && \mathrm{exp}\left(h(A +_2 B)\right) =& \mathrm{exp}\left(h(A) + h(B)\right)
        = \mathrm{exp}(h(A))\cdot\mathrm{exp}(h(B));\\
        \mathrm{(iii)} && \mathrm{exp}\left(h(\mathbb{O}_N)\right) =& \mathbb{I}_{N^2},
        \quad \mathrm{cexp}\left(h(\mathbb{O}_N)\right) = \mathbb{I}_{2N^2};\,
    \end{align}
    \end{subequations}
    where $\mathbb{O}_N$ is the $N \times N$ matrix of all zeros. 
    
    In addition, statements (i) and (ii) continue to hold if $\mathrm{exp}(h(\cdot))$ is substituted with
    $\mathrm{cexp}(h(\cdot))$.
\end{lemma}
\begin{proof}
    We start from claim (i). For all $A\in\mysymmsetztwo{N}$, we know that $h(A)$ is skew-symmetric and diagonal, so it follows that $h(A)^*=-h(A)$, but $h(A)^\top=h(A)$, so
    \begin{equation}
        \label{hprops}
        h(A)^*=\overline{h(A)^\top}=\overline{h(A)}=-h(A)\,.
    \end{equation}
    Again, since $h(A)$ is diagonal and the complex conjugate is commutative under composition with the exponential mapping, by applying
    $\mathrm{exp}$ to \eqref{hprops} we obtain the statement in \eqref{inside1}.

    For claim (ii), it can be shown by exhaustion that for all $a, b\in\mathbb{Z}_2$, $(-1)^{a +_2 b}=(-1)^{a+b}$, that is 
    $e^{\imath\pi(a +_2 b)}=e^{\imath\pi(a+b)}$.
    Thus,
    \begin{subequations}
    \begin{align}
         \mathrm{exp}\left(h(A +_2 B)\right) &= \mathrm{exp}\left(\imath\pi \sum_{i, j} (A_{i, j} +_2 B_{i, j}) \myouter{i, j}{i, j}\right)\\
         &= \mathrm{exp}\left(\imath\pi \sum_{i, j} (A_{i, j} + B_{i, j}) \myouter{i, j}{i, j}\right) =  \mathrm{exp}\left(h(A) + h(B)\right)\,.
    \end{align}
    \end{subequations}
    Now, $h(A)$ and $h(B)$ are diagonal, so their product commute,
    then from the Campbell-Baker-Hausdorff series \cite{rossmann2006lie}, with commutator $[h(A), h(B)]=0$
    it follows that $\mathrm{exp}(h(A)+h(B))=\mathrm{exp}(h(A))\cdot\mathrm{exp}(h(B))$.

    The additional claim regarding the controlled variant of (i) and (ii), is a consequence of \eqref{cnot_decomp} and of the fact that the
    controlled operator is itself diagonal. We omit the details for this case.

    Finally for claim (iii), we have that
    \begin{subequations}
    \begin{align}
        \mathrm{exp}\left(h(\mathbb{O}_N)\right) =& \mathrm{exp}\left(\imath\pi \sum_{i, j} 0 \myouter{i, j}{i, j}\right) = \sum_{i, j} \myouter{i, j}{i, j} = \mathbb{I}_{N^2}\,,\\
        \mathrm{cexp}\left(h(\mathbb{O}_N)\right)
        =& \mathbb{I}_{N^2} \oplus \mathbb{I}_{N^2} = \mathbb{I}_{2N^2}\,.
    \end{align}
    \end{subequations}
\end{proof}

The image of the map $h$ is a subset of the Lie algebra\cite{rossmann2006lie} $\mathfrak{u}\left(N^2\right)$, which consists of a set of $N^2 \times N^2$
skew-hermitian matrices with some additional structure not relevant for the present work. 
We can then write $h(\mathcal{S}_N(\mathbb{Z}_2)) \subset \mathfrak{u}\left(N^2\right)$,
and therefore, the matrix exponential map takes $h(\mathcal{S}_N(\mathbb{Z}_2))$ to the \textit{unitary group} of degree $N^2$, denoted by $\mathsf{U}(N^2)$.
By using the latter fact we introduce, for the adjacency matrix $A \in \mathcal{S}_N(\mathbb{Z}_2)$, a
\textit{unitary operator representation} denoted $\widehat{A} \in \mathsf{U}(2N^2)$ and defined as
\begin{subequations}
\begin{align}
    \label{adj_hat_def}
    \widehat{A} \coloneqq& \mathrm{Ad}\left(H^{\otimes (2k+1)}\right) \mathrm{cexp}(h(A))\\
    =& \mathrm{Ad}\left(H^{\otimes (2k+1)}\right) \left(\mathbb{I}_{N^2} \oplus \mathrm{exp}(h(A))\right)\\
    \label{adj_hat_def:cexp_expanded}
    =& H^{\otimes (2k+1)} \left(\mathbb{I}_{N^2} \oplus \sum_{i, j} (-1)^{A_{i, j}} \myouter{i, j}{i, j}\right) H^{\otimes (2k+1)}
\end{align}
\end{subequations}
where $H$ is the $2\times 2$ matrix corresponding to the Hadamard operator on a single qubit. 

We note here that the representation $\widehat{A} \in \mathsf{U}(2N^2)$ requires a quantum circuit of $2k + 1$ qubits, where $k=\log_2(N)$ by construction.
In this sense, $\widehat{A}$ represents the adjacency $A \in \mathcal{S}_N(\mathbb{Z}_2)$ using a number of qubits that is logarithmic in the number of nodes in the graph,
and this fact is one of the cornerstones of our contribution. We will refer to the representation $\widehat{A}$ as the \emph{log-Hadamard representation} of the adjacency
$A$ of graph $\mathcal{A}$. While at this point the reason beyond this specific form for $\widehat{A}$ (besides its logarithmic scaling) could seem unclear, they will be elucidated in the following.  

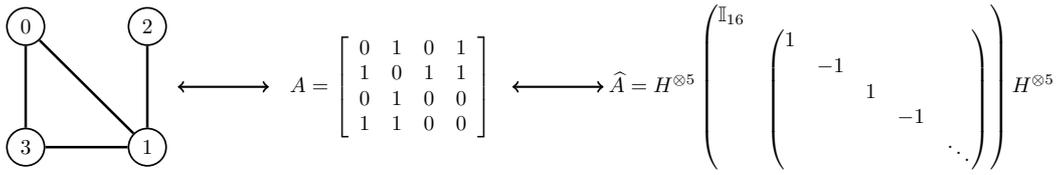
\begin{figure}
    \centering \vspace*{0.5cm}
    \scalebox{0.8}{
    \begin{tikzpicture}
        \begin{scope}[every node/.style={circle,thick,draw}]
            \node (3) at (0,0) {3};
            \node (0) at (0,2) {0};
            \node (2) at (2,2) {2};
            \node (1) at (2,0) {1};
        \end{scope}
        \begin{scope}[
            every node/.style={fill=white,circle},
            every edge/.style={draw=black,very thick}]
            \path [-] (0) edge (1);
            \path [-] (0) edge (3);
            \path [-] (1) edge (3);
            \path [-] (1) edge (2);
        \end{scope} 
        \draw[arrows=<->,line width=1pt](2.5,1)--(4,1);
        \node at (6.0,1) {$A = \left[\begin{array}{cccc} 
        0 & 1 & 0 & 1 \\ 
        1 & 0 & 1 & 1 \\
        0 & 1 & 0 & 0 \\
        1 & 1 & 0 & 0  \end{array}\right]$};
        \draw[arrows=<->,line width=1pt](8.0,1)--(9.5,1);
        \node at (13.25,1) {$\widehat{A} = H^{\otimes5}
        \begin{pmatrix}\mathbb{I}_{16}&\\&
            \begin{pmatrix}
                1 &\\
                & -1\\
                & & 1\\
                & & & -1\\
                & & & &\ddots
            \end{pmatrix}\end{pmatrix}
         H^{\otimes5} $};
    \end{tikzpicture}}
    \caption{A sample graph for demonstrating the log-Hadamard representation.}
    \label{first_sample_graph}
\end{figure}

We report in Figure~\ref{first_sample_graph} and example of a graph, its adjacency matrix, and its log-Hadamard representation. 

\begin{remark}
\rev{We cite here the very recent work~\cite{Rancic2021}, where a different yet similar-in-spirit transformation is used as a starting point to tackle the max-cut problem with a logarithmic scaling.}
\end{remark}

\subsection{Circuit construction for the log-Hadamard representation}

We now look briefly at how to encode the log-Hadamard representation $\widehat{A}$ as a quantum circuit featuring two-qubit gates, and in particular one can show that one has multiple options. First, we notice that the circuit construction for $\widehat{A}$ is obtained through conjugation by the Walsh-Hadamard transform acting on a diagonal operator
\cite{Shende_2006} depending on the $A_{i, j}$, so it is something well understood. As expected, the representation requires $\log_2\left(2^k\times 2^k\right)+1=2k+1$ qubits.

Different approaches then exist to compile such operator for two-qubit gates in hardware. A direct strategy is reported in Figure~\ref{simple_circ_for_adj} for the $4$-vertex graph reported in Figure~\ref{first_sample_graph} in terms of $X$, $H$ and multi-controlled $Z$ gates. In essence, the construction consists on the decomposition of the diagonal, having values in $\{1, -1\}$, as composition of simpler diagonals
each having a single element $-1$. The number of layers (multi-controlled phase gate and $X$ gates) depends linearly on the number of edges of the represented graph. 

\begin{figure}
    \centering
    \begin{align*}
        \scalebox{0.6}{
        \Qcircuit @C=0.5em @R=0.25em @!R {
            & \lstick{ {j}_{0} :  } & \gate{\mathrm{H}} & \gate{\mathrm{U_1}\,(\mathrm{\frac{\pi}{8}})} & \targ & \targ & \gate{\mathrm{U_1}\,(\mathrm{\frac{\pi}{8}})} & \targ & \gate{\mathrm{U_1}\,(\mathrm{\frac{\pi}{4}})} & \targ & \gate{\mathrm{U_1}\,(\mathrm{\frac{\pi}{8}})} & \targ & \targ & \gate{\mathrm{U_1}\,(\mathrm{-\frac{-\pi}{8}})} & \targ & \targ & \targ & \gate{\mathrm{U_1}\,(\mathrm{\frac{\pi}{8}})} & \targ & \targ & \gate{\mathrm{U_1}\,(\mathrm{-\frac{-\pi}{8}})} & \targ & \gate{\mathrm{U_1}\,(\mathrm{-\frac{-\pi}{4}})} & \targ & \gate{\mathrm{U_1}\,(\mathrm{-\frac{-\pi}{8}})} & \targ & \targ  & \qw \\
            & \lstick{ {j}_{1} :  } & \gate{\mathrm{H}} & \qw & \ctrl{-1} & \qw & \qw & \ctrl{-1} & \qw & \qw & \qw & \ctrl{-1} & \qw & \qw & \ctrl{-1} & \qw & \ctrl{-1} & \qw & \qw & \ctrl{-1} & \qw & \qw & \qw & \ctrl{-1} & \qw & \qw & \ctrl{-1}  & \qw \\
            & \lstick{ {i}_{0} :  } & \gate{\mathrm{H}} & \qw & \qw & \ctrl{-2} & \qw & \qw & \qw & \qw & \qw & \qw & \ctrl{-2} & \qw & \qw & \qw & \qw & \qw & \ctrl{-2} & \qw & \qw & \qw & \qw & \qw & \qw & \ctrl{-2} & \qw  & \qw \\
            & \lstick{ {i}_{1} :  } & \gate{\mathrm{H}} & \qw & \qw & \qw & \qw & \qw & \qw & \ctrl{-3} & \qw & \qw & \qw & \qw & \qw & \qw & \qw & \qw & \qw & \qw & \qw & \ctrl{-3} & \qw & \qw & \qw & \qw & \qw  & \qw\\
            & \lstick{ {b}     :  } & \gate{\mathrm{H}} & \qw & \qw & \qw & \qw & \qw & \qw & \qw & \qw & \qw & \qw & \qw & \qw & \ctrl{-4} & \qw & \qw & \qw & \qw & \qw & \qw & \qw & \qw & \qw & \qw & \qw  & \qw
        }}\\
        \scalebox{0.6}{
        \Qcircuit @C=0.5em @R=0.25em @!R {
            & \gate{\mathrm{U_1}\,(\mathrm{-\frac{-\pi}{8}})} & \qw & \targ & \qw & \qw & \qw & \qw & \qw & \qw & \qw & \qw & \qw & \qw & \qw & \qw & \qw & \qw & \qw & \qw & \qw & \qw & \qw & \qw & \qw & \qw & \gate{\mathrm{H}} & \qw & \qw\\ 
            & \gate{\mathrm{U_1}\,(\mathrm{-\frac{-\pi}{8}})} & \targ & \qw & \targ & \gate{\mathrm{U_1}\,(\mathrm{-\frac{-\pi}{8}})} & \targ & \gate{\mathrm{U_1}\,(\mathrm{\frac{\pi}{4}})} & \targ & \gate{\mathrm{U_1}\,(\mathrm{-\frac{-\pi}{4}})} & \targ & \gate{\mathrm{U_1}\,(\mathrm{\frac{\pi}{8}})} & \targ & \targ & \gate{\mathrm{U_1}\,(\mathrm{\frac{\pi}{8}})} & \qw & \targ & \qw & \qw & \qw & \qw & \qw & \qw & \qw & \qw & \qw & \gate{\mathrm{H}} & \qw & \qw\\ 
            & \qw& \ctrl{-1} & \qw & \qw & \qw & \ctrl{-1} & \qw & \qw & \qw & \ctrl{-1} & \qw & \qw & \ctrl{-1} & \gate{\mathrm{U_1}\,(\mathrm{\frac{\pi}{8}})} & \targ & \qw & \targ & \targ & \gate{\mathrm{U_1}\,(\mathrm{-\frac{-\pi}{8}})} & \targ & \qw & \qw & \qw & \qw & \qw & \gate{\mathrm{H}} & \qw & \qw\\
            & \qw& \qw & \qw & \ctrl{-2} & \qw & \qw & \qw & \qw & \qw & \qw & \qw & \ctrl{-2} & \qw & \qw & \ctrl{-1} & \qw & \qw & \ctrl{-1} & \gate{\mathrm{U_1}\,(\mathrm{-\frac{-\pi}{8}})} & \qw & \targ & \gate{\mathrm{U_1}\,(\mathrm{\frac{\pi}{8}})} & \targ & \qw & \qw & \gate{\mathrm{H}} & \qw & \qw\\ 
            & \qw& \qw & \ctrl{-4} & \qw & \qw & \qw & \qw & \ctrl{-3} & \qw & \qw & \qw & \qw & \qw & \qw & \qw & \ctrl{-3} & \ctrl{-2} & \qw & \qw & \ctrl{-2} & \ctrl{-1} & \qw & \ctrl{-1} & \gate{\mathrm{U_1}\,(\mathrm{\frac{\pi}{2}})} & \qw & \gate{\mathrm{H}} & \qw & \qw
        }}
    \end{align*}
    \caption[Qiskit circuit implementation]{An implementation of the circuit (using the Diagonal gate of Qiskit) for $\widehat{A}$ where $A$
    is the adjacency matrix for the graph depicted in Figure~\ref{first_sample_graph}. The gate $U_1$ is a synonym of the phase gate with
    $U_1(\lambda)=\big(\begin{smallmatrix} 1 & 0 \\ 0 & e^{\imath\lambda} \end{smallmatrix}\big)$.}
    \label{qiskit_circ_for_adj}
\end{figure}
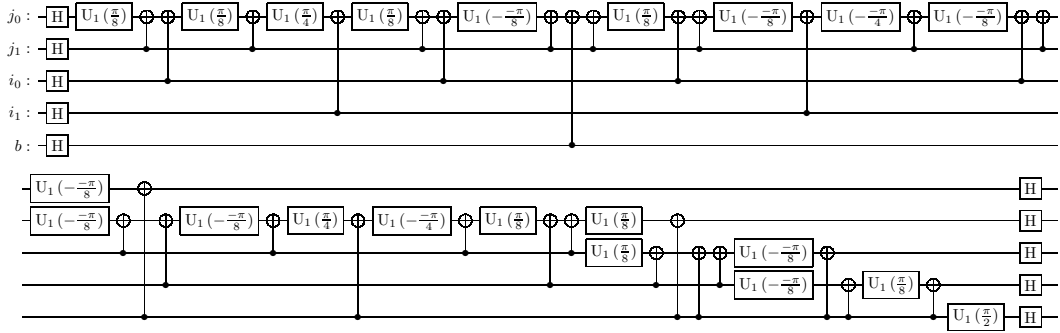

In Figure~\ref{qiskit_circ_for_adj}, we report instead a second strategy, which is the one obtained by means of the diagonal operator implemented in the IBM Qiskit software \cite{Qiskit}. At the time of writing the latter implementation is based on the decomposition of diagonal operators as explained in Theorem 7 of \cite{Shende_2006}. The proposed decomposition relies on Theorem 8 of \cite{Shende_2006} concerning the demultiplexing of multiplexed $R_k$ gates (with $k \in \{\sigma_y, \sigma_z\})$),
also known as \textit{uniformly controlled rotations}. As reported in the same theorem, such demultiplexing results in an exponential number of CNOTs with respect to the number of
controlling qubits.
In our case the number of controlling qubits is $2k$ for a graph of $N=2^k$ vertices, hence the number of CNOTs for the hat representation has complexity $O(N^2)$.
In regard to the latter complexity, $O\left(N^2\right)=O\left(2^{2k}\right)$, we could compare it to what is needed in QUBO approaches that require $q$ qubits (either $q=O(N^2)$ or $q=O(N \lceil \log_2 N \rceil)$), and need a worst-case depth of $O(q^2 p)$, with $p$ the depth of the employed QAOA solver. Moreover, $O(N^2)$ needs to be also compared to the search space of the problem we are approaching, which is $N!$. 

Finally, we leave for future research more wholistic approaches that could take the problem circuit and compile it in an approximate way to smaller depths (a topic of growing research attention){~\cite{khatri2019, younis2021qfast, Madden2021}}.

\begin{figure}
    \centering
    \begin{equation*}
        \scalebox{0.7}{
        \Qcircuit @C=0.5em @R=0.25em @!R {
            & \lstick{ {j}_{0} :} & \gate{\mathrm{H}} & \qw               & \ctrl{1}     & \qw               & \ctrl{1}     & \gate{\mathrm{X}} & \ctrl{1}     & \qw               & \ctrl{1}     & \gate{\mathrm{X}} & \ctrl{1}     & \qw               & \ctrl{1}     & \gate{\mathrm{X}} & \ctrl{1}     & \gate{\mathrm{X}} & \ctrl{1}     & \qw               & \gate{\mathrm{H}} & \qw\\ 
            & \lstick{ {j}_{1} :} & \gate{\mathrm{H}} & \gate{\mathrm{X}} & \ctrl{1}     & \gate{\mathrm{X}} & \ctrl{1}     & \gate{\mathrm{X}} & \ctrl{1}     & \gate{\mathrm{X}} & \ctrl{1}     & \qw               & \ctrl{1}     & \gate{\mathrm{X}} & \ctrl{1}     & \qw               & \ctrl{1}     & \qw               & \ctrl{1}     & \gate{\mathrm{X}} & \gate{\mathrm{H}} & \qw\\ 
            & \lstick{ {i}_{0} :} & \gate{\mathrm{H}} & \gate{\mathrm{X}} & \ctrl{1}     & \qw               & \ctrl{1}     & \gate{\mathrm{X}} & \ctrl{1}     & \qw               & \ctrl{1}     & \qw               & \ctrl{1}     & \gate{\mathrm{X}} & \ctrl{1}     & \gate{\mathrm{X}} & \ctrl{1}     & \qw               & \ctrl{1}     & \qw               & \gate{\mathrm{H}} & \qw\\ 
            & \lstick{ {i}_{1} :} & \gate{\mathrm{H}} & \gate{\mathrm{X}} & \ctrl{1}     & \qw               & \ctrl{1}     & \qw               & \ctrl{1}     & \qw               & \ctrl{1}     & \qw               & \ctrl{1}     & \gate{\mathrm{X}} & \ctrl{1}     & \qw               & \ctrl{1}     & \qw               & \ctrl{1}     & \qw               & \gate{\mathrm{H}} & \qw\\ 
            & \lstick{ {b} :}     & \gate{\mathrm{H}} & \qw               & \control \qw & \qw               & \control \qw & \qw               & \control \qw & \qw               & \control \qw & \qw               & \control \qw & \qw               & \control \qw & \qw               & \control \qw & \qw               & \control \qw & \qw               & \gate{\mathrm{H}} & \qw\\ 
        }}
    \end{equation*}
    \caption{A direct implementation of the circuit for $\widehat{A}$ where $A$ is the adjacency matrix for the
    graph depicted in Figure \ref{first_sample_graph}.
    Here the qubit labels $i_t$ and $j_t$ (where $j_0$ is the least significative qubit) correspond to the qubits representing the index of the entries of the adjacency matrices
    according to the definition \eqref{adj_hat_def:cexp_expanded}. The qubit $b$ is the most significant one and corresponds to the control for the $\mathrm{cexp}$.}
    \label{simple_circ_for_adj}
\end{figure}
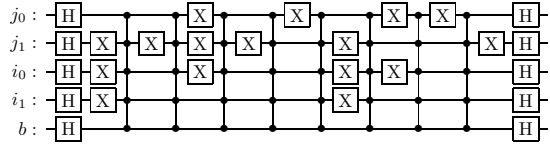

\subsection{Operations on the representation}

Given the log-Hadamard representation for the adjacency matrix, we are now ready to build operations that can be carried over the representation, and which will be key in building the loss function. 
First of all, we consider addition (and subtraction) between adjacency matrices $A$ and $B$ of the same size, which is captured by the following proposition. 

\begin{theorem}
    \label{hataplusb}
    The log-Hadamard representation $\widehat{(\cdot)}$ is a monomorphism\footnote{Injective group-homomorphism} between the additive group on the set $\mathcal{S}_N(\mathbb{Z}_2)$
    and the corresponding group of unitary operators under composition. That is, for all $A, B \in \mathcal{S}_N(\mathbb{Z}_2)$, we have:
\begin{eqnarray*}
\mathrm{(i)} & \widehat{\left(A +_2 B\right)} = \widehat{A}\cdot \widehat{B} = \widehat{B}\cdot\widehat{A},\quad \widehat{\mathbb{O}_N}=\mathbb{I}_{2N^2} & \text{ (group-homomorphism) }\\
\mathrm{(ii)} & A\ne B \implies \widehat{A} \ne \widehat{B} & \text{ (injectivity) } 
\end{eqnarray*}

Finally, the property that $x\in \mathbb{Z}_2 \implies x=(-x)$ is reflected into the fact that $\widehat{(-A)}=(\widehat{A})^{*}=\widehat{A}$
for all $A \in \mathcal{S}_N(\mathbb{Z}_2)$, hence $\widehat{\left(A +_2 (-B)\right)} = \widehat{\left(A +_2 B\right)}$.
\end{theorem}

\begin{proof}
    By Lemma \ref{lemma:exph} and Eq.~\eqref{ad_homom}, we have
    \begin{subequations}
    \begin{align}
        \widehat{\left(A +_2 B\right)} =& \myhatexp{h\left(A +_2 B\right)}\\
        =& \myhatexp{h(A) + h(B)}\\
        =& \myhatexp{h(A)} \cdot \myhatexp{h(B)} = \widehat{A} \cdot \widehat{B}\,,
    \end{align}
    \end{subequations}
    also since $\mathrm{cexp}(h(A))$ is diagonal, then it is clear that the image of $\widehat{(\cdot)}$ is a subgroup of operators sharing the same eigenvectors,
    consequently $\widehat{A}$ and $\widehat{B}$ commute, that is the abelian property is preserved, hence $\widehat{A}\cdot \widehat{B} = \widehat{B}\cdot\widehat{A}$.
    Moverover,
    \begin{align}
        \widehat{\mathbb{O}_N} =& \myhatexp{h\left(\mathbb{O}_N\right)}= \myadh{(2k+1)} \mathbb{I}_{2N^2} = \mathbb{I}_{2N^2}\,,
    \end{align}
    thus $\widehat{(\cdot)}$ is a group-homomorphism.
    We now prove the contrapositive of \emph{(ii)}. Assume $\widehat{A}=\widehat{B}$, then
    \begin{subequations}
    \begin{align}
         & \myhatexp{h(A)} = \myhatexp{h(B)}\\
         \implies& \mathrm{exp}\, h(A) = \mathrm{exp}\, h(B)\\
         \implies& \sum_{i, j} (-1)^{A_{i, j}} \myouter{i, j}{i, j} =
         \sum_{i^{\prime}, j^{\prime}} (-1)^{B_{i^{\prime}, j^{\prime}}} \myouter{i^{\prime}, j^{\prime}}{i^{\prime}, j^{\prime}}\\
         \implies& (-1)^{A_{i, j}} = (-1)^{B_{i, j}}\quad\forall\,i,j
    \end{align}
    \end{subequations}
    but $A_{i, j}, B_{i, j} \in \mathbb{Z}_2$, then $A_{i, j}=B_{i, j}$, that is $\widehat{A}=\widehat{B} \implies A=B$ so \emph{(ii)} is proven by contrapositive;
    hence the group-homomorphism is injective.

    Finally, we prove the self-adjointness of $\widehat{A}$, that is $\left(\widehat{A}\right)^{*}=\widehat{A}$.
    By Lemma \ref{lemma:exph}, we have that $\mathrm{cexp}\left(h(A)\right) = \mathrm{cexp}\left(h(A)\right)^*$, then,
    since $H=H^*$, we conclude that
    \begin{subequations}
    \begin{align}
        & \mathrm{cexp}\left(h(A)\right) = \mathrm{cexp}\left(h(A)\right)^*\\
        \iff& \myadh{(2k+1)}\mathrm{cexp}\left(h(A)\right) = \myadh{(2k+1)}\mathrm{cexp}\left(h(A)\right)^*\\
        \iff& \widehat{A} = \left(\widehat{A}\right)^*\,.
    \end{align}
    \end{subequations}
\end{proof}

Proposition~\ref{hataplusb} is key in our construction, since it provides a way to build our loss and translating classical operations into ``hat'' operations.
One of the important facts is the injectivity, such that, from a unitary perspective, if $\widehat{A} = \widehat{B}$ then $A = B$, which is key in building a loss
on $\widehat{A}, \widehat{B}$ to measure something on $A, B$. 

However, when talking about physical systems, we do not just care about unitary operations, but also physical meaning. In particular, we may wonder if it can happen that $\widehat{A}$ and $\widehat{B}$ are different, yet physically indistinguishable, because e.g., they are the same up to a global phase. In practice this would mean that there could be cases in which $A, B$ are different, thus $\widehat{A}$ and $\widehat{B}$ are different, but the latter is not physically detectable and any physical function would assume that $\widehat{A}$ and $\widehat{B}$ are actually the same (again, think about global phases).

Next we show that this is not possible for adjacency matrices, ultimately proving that our approach can work in practice. 

\subsection{Physical \rev{distinguishability}}
\label{subsection:phy_distinguishability}

As said, we have concerns about the physical \rev{distinguishability} of the quantum state produced by the operator $\widehat{(.)}$, that is whether there exist
adjacency matrices $A, B \in \mysymmsetztwo{N}$, such that $A\ne B$ and their corresponding representations $\widehat{A}$ and $\widehat{B}$ cannot be physically distinguished. We show that this is not the case by introducing the `doubling' functor \cite{SELINGER2007139} $\mathcal{D}$, which in our context suffices to be interpreted as a mapping
having as domain the unitary group $\mathsf{U}(2N^2)$ and rule $f \mapsto \overline{f}\otimes f$.
The codomain of $\mathcal{D}$ is still a unitary group with identity $\mathbb{I}_{2N^2} \otimes \mathbb{I}_{2N^2}$.

The convenience of the `doubling' $\mathcal{D}$ can be appreciated by considering a unitary operator $f$ and an arbitrary global phase $e^{\imath\theta}$ for some
$\theta \in \mathbb{R}$, then
\begin{align}
    \mathcal{D}\left(e^{\imath\theta}f\right) = \overline{\left(e^{\imath\theta}f\right)} \otimes \left(e^{\imath\theta}f\right)= \left(e^{-\imath\theta}\overline{f}\right) \otimes \left(e^{\imath\theta}f\right)= \overline{f} \otimes f = \mathcal{D}(f),
\end{align}
that is, for all vectors $v$ belonging to the domain of $f$, the evaluation of $e^{\imath\theta}f$ and $f$ at $v$, 
gives states that are physically identical for any $\theta \in \mathbb{R}$.
Also `doubling' preserves commutativity w.r.t. operator composition, indeed for any unitaries $U_1$ and $U_2$ acting on the same vector space, such that
$U_1 U_2=U_2 U_1$, we have
\begin{subequations}
\label{doubling_pres_abel}
\begin{align}
    \mathcal{D}(U_1 U_2)=&\overline{U_1 U_2} \otimes U_1 U_2
    =\left(\overline{U_1} \otimes U_1\right)\cdot \left(\overline{U_2} \otimes U_2\right)
    =\mathcal{D}(U_1)\cdot \mathcal{D}(U_2)\\
    =&\overline{U_2 U_1} \otimes U_2 U_1
    =\left(\overline{U_2} \otimes U_2\right)\cdot \left(\overline{U_1} \otimes U_1\right)
    =\mathcal{D}(U_2)\cdot \mathcal{D}(U_1)\,.
\end{align}
\end{subequations}
In this context, the $\mathcal{D}$ serves the purpose of detecting whether any two unitary operators can be distinguished: if their composition through $\mathcal{D}$ outputs the same quantity, then they are indistinguishable. 

Let $\mathbb{O}_N$ denote the $N\times N$ matrices of zeros to be considered  belonging to $\mysymmsetztwo{N}$, which is in fact the identity element w.r.t the group on $\mathcal{S}_N(\mathbb{Z}_2)$.
Then the following result is in place, the proof is presented in the Appendix \ref{appendix_more_proofs}.

\begin{theorem}
    \label{dbl_hat_not_faithful_proof}
    Given the $N\times N$ adjacency matrices $A, B \in \mathcal{S}_N(\mathbb{Z}_2)$, 
    the functor composition $\mathcal{D}(\widehat{(\cdot)})$  is a group-homomorphism, that is
    \begin{align}
        \mathcal{D}\left(\widehat{A +_2 B}\right) =& \mathcal{D}\left(\widehat{A}\right) \cdot \mathcal{D}\left(\widehat{B}\right)
        =\mathcal{D}\left(\widehat{B}\right) \cdot \mathcal{D}\left(\widehat{A}\right)\,,\\
        \quad\mathcal{D}\left(\widehat{\mathbb{O}_N}\right) =& \mathbb{I}_{2N^2} \otimes \mathbb{I}_{2N^2}\,.
    \end{align}
    In addition, the kernel of the composition is trivial, and in particular $\mathrm{ker}\, \mathcal{D}\left(\,\widehat{(\cdot)}\,\right) = \{\mathbb{O}_N\}$, consequently the homomorphism is injective.
\end{theorem}

Proposition~\ref{dbl_hat_not_faithful_proof} says that when $\widehat{A}$ and $\widehat{B}$ are different, they are also physically distinguishable,
since the kernel of $\mathcal{D}(\widehat{(\cdot)})$ is the identity element of $A$. Therefore, whenever $A\neq B$ a loss built on $\widehat{A}$ and
$\widehat{B}$ will be able to tell.
By taking the contrapositive, if from the physical circuit we observe that $\widehat{A} = \widehat{B}$, then $A = B$. \rev{Proposition~\ref{dbl_hat_not_faithful_proof} show that we can use $\widehat{A}, \widehat{B}$ instead of $A$, $B$ to build our cost}\footnote{\rev{Note that the `doubling' will not be implemented in the final circuit, it is only a way to show the correctness of our approach.}}.  

\begin{remark}
    A remark here is in order to highlight the need for $\mathrm{cexp}$ in the definition of $\widehat{(\cdot)}$. If one had set
    $\widehat{A}=\mathrm{Ad}\left(H^{\otimes 2k}\right) \mathrm{exp}(h(A))$, it could be shown that
    $\mathrm{ker}\, \mathcal{D}\left(\,\widehat{(\cdot)}\,\right) \neq \{\mathbb{O}_N\}$,
    which implies that we would loose the physical distinguishability. Moreover, this is not the only advantage, indeed in the construction of the loss we will highlight the presence of a metric structure. \hfill $\Box$
\end{remark}

\subsection{Permutation of the log-Hadamard representation}

All the elements are now in place to introduce the permutation of the log-Hadamard representation, which is the last step to be able to formulate the sub-graph isomorphism problem as a quantum computing problem. We start by reminding that, as for the classical loss, we have looked at
\begin{equation}
    \ell_{\mathrm{C}}(P; A, B) = \left\|S P A P^\top S^{\top} - B \right\|^2_{\mathrm{F}}, 
\end{equation}
 where $P\in \Pi_N$ is a permutation matrix and $S$ is the selection matrix.

In this section, we show how the permutation $P$ acting classically `under the hat', that is $\widehat{PAP^{\top}}$, can be moved out of the
operator $\widehat{(\cdot)}$ to be implemented in terms of quantum gates.
The aim is that of obtaining a parametric quantum circuit $\tilde{P}(\mathbf{\theta})$ for the permutation $P$ that acts 
on the adjacency representation $\widehat{A}$. The parameters of $\tilde{P}$ can then be tuned by a classical optimizer in a process
that follows the variational approach scheme. In a second time, we will see how to take care of the selection $S$.

We start by reintroducing the parametrization of $P$ as the result of a permutation function $p$ on the vertices of the graph and the induced group-homomorphism as in Eq.~\eqref{p-P}.  Then the following result holds. 

\begin{theorem}\label{th-PAP}
    Let $\bar{p}\in\mathsf{S}_N$ be an element of the symmetric group and $P_{\bar{p}}$ the corresponding permutation matrix as $P_{\bar{p}}= \sum_{i=0}^{N-1}\ket{\bar{p}(i)}\bra{i}$. Define the representation for the permutation $P_{\bar{p}}$ as
    \begin{align}
        \label{hat_for_p}
        \check{P}_{\bar{p}}=&\myiden{2} \otimes \left(H^{\otimes 2k}\cdot P_{{\bar{p}}}^{\otimes 2} \cdot H^{\otimes 2k}\right)
    \end{align}
    then the conjugate permutation under the hat corresponds to the following composition of representations
    \begin{equation}\label{eq.PAP}
        \widehat{P_{\bar{p}}AP_{\bar{p}}^{\top}} = \check{P}_{\bar{p}}\cdot \widehat{A}\cdot \check{P}_{\bar{p}}^{\top}.
    \end{equation}
\end{theorem}

\begin{proof} 
	We note that the composition of $\check{(\cdot)}$ and $(\cdot)^\top$ commutes, and therefore $\left(\check{P}\right)^\top=\widecheck{\left(P^\top\right)}$.
    Then, we expand $P_{{\bar{p}}}AP_{{\bar{p}}}^{\top}$ as follows
    \begin{align}
        P_{{\bar{p}}}AP_{{\bar{p}}}^{\top} =& \left(\sum_{i^{\prime}=0}^{N-1}\ket{{\bar{p}}(i^{\prime})}\bra{i^{\prime}}\right)\cdot
        \left(\sum_{i,j}A_{i, j}\ket{i}\bra{j}\right)\cdot
        \left(\sum_{j^{\prime}=0}^{N-1} \ket{j^{\prime}} \bra{{\bar{p}}(j^{\prime})}\right)= \sum_{i,j}A_{i, j}\ket{{\bar{p}}(i)}\bra{{\bar{p}}(j)}\,,
    \end{align}
    also note that, by considering the outer product expansion of $\mathrm{cexp}$ in \eqref{adj_hat_def:cexp_expanded}, we obtain
    \begin{subequations}
    \begin{align}
        \mathrm{cexp}\left(h\left(P_{{\bar{p}}}AP_{{\bar{p}}}^{\top}\right)\right) =& 
        \myiden{N^2} \oplus \sum_{i, j} (-1)^{A_{i,j}} \ket{{\bar{p}}(i), {\bar{p}}(j)} \bra{{\bar{p}}(i), {\bar{p}}(j)}\\
        =&\myiden{N^2} \oplus \sum_{i, j} (-1)^{A_{i,j}} P_{{\bar{p}}}^{\otimes 2} \myouter{i, j}{i, j} \left(P_{{\bar{p}}}^\top\right)^{\otimes 2} \\
        =& 
        \myiden{N^2} \oplus \Big( P_{{\bar{p}}}^{\otimes 2} \Big[\sum_{i, j} (-1)^{A_{i,j}}  \myouter{i, j}{i, j} \Big]\left(P_{{\bar{p}}}^\top\right)^{\otimes 2} \Big)
        \\
        \underbrace{=}_{\text{Eq.~\eqref{cnot_conj_decomp}}}& \left(\myiden{2} \otimes P_{{\bar{p}}}^{\otimes 2}\right)
        \cdot \mathrm{cexp}\left(h(A)\right)
        \cdot \left(\myiden{2} \otimes \left(P_{{\bar{p}}}^{\top}\right)^{\otimes 2}\right)\,.
    \end{align}
    \end{subequations}
    Then, by applying the log-Hadamard representation $\widehat{(\cdot)}$ to the matrix
    $P_{{\bar{p}}}AP_{{\bar{p}}}^{\top} \in \mathcal{S}_{N}(\mathbb{Z}_2)$, we get
    \begin{subequations}
    \begin{align}
        \widehat{P_{{\bar{p}}}AP_{{\bar{p}}}^{\top}} =& 
        H^{\otimes (2k+1)}
        \cdot \mathrm{cexp}\left(h\left(P_{{\bar{p}}}AP_{{\bar{p}}}^{\top}\right)\right)
        \cdot H^{\otimes (2k+1)}\\
        =& H^{\otimes (2k+1)}\cdot \left(\myiden{2} \otimes P_{{\bar{p}}}^{\otimes 2}\right)
        \cdot \mathrm{cexp}\left(h(A)\right)
        \cdot \left(\myiden{2} \otimes \left(P_{{\bar{p}}}^{\top}\right)^{\otimes 2}\right) \cdot H^{\otimes (2k+1)}\\
        =& \check{P}_{\bar{p}}\cdot \widehat{A}\cdot \check{P}_{\bar{p}}^{\top}
    \end{align}
    \end{subequations}
    as required.
\end{proof}

Proposition~\ref{th-PAP} tells us how $P A P^{\top}$ get transformed under the log-Hadamard representation and allows us to devise a quantum circuit for it. In particular, the key ingredients are Eq.~\eqref{hat_for_p} and Eq.~\eqref{eq.PAP}. The former explains how $P$ is affected, while the latter tells us how $P A P^{\top}$ is transformed. 

Looking at both equations, given the unitary representation of the adjacency matrix $A$ as $\widehat{A}$, the action of the representation
for the permutation $\check{P}_{\bar{p}}$, corresponds to the same action of the permutation $P_{\bar{p}}$ in the classical sense. 
The decomposition in \eqref{eq.PAP} can be interpreted, on the left-hand side of the equation, as the permutation applied
classically prior to the transformation into unitary representation. The right-hand side of \eqref{eq.PAP} instead,
shows a permutation applied in the quantum domain (after the hat operator).
The representation for the adjacency matrix permutation operator in \eqref{hat_for_p} shows that the circuit implementing the
mechanism will present two identical Ansatze stacked to form the equivalent of $P_{\bar{p}} \otimes P_{\bar{p}} = P_{\bar{p}}^{\otimes 2}$, which is quite interesting. We remark that $P_{\bar{p}} \in \Pi_N$, so $P_{\bar{p}}^{\otimes 2}$ is a $N^2 \times N^2$ matrix and has the same dimensions of $H^{\otimes 2k}$. 

Finally, in \eqref{eq.PAP} we see that the stacked Ansatze appear pre- and post-composed to $\widehat{A}$, however,
below we will show that the expression, when part of the loss, simplifies and the post-composition cancels out.

\subsection{The quantum loss function}
\subsubsection{Graph isomorphism case}
We now look at how to build the quantum loss function to represent the classical loss $\ell_{\mathrm{C}}$ into the quantum domain. To facilitate the exposition, we start with graphs of the same size, and then we extend it to our sub-graph isomorphism problem, featuring graphs of different sizes. Consider then the adjacency matrices $A, B \in \mathcal{S}_N(\mathbb{Z}_2)$.

We start by defining a classical disparity function $\lossc: \mysymmsetztwo{N} \times \mysymmsetztwo{N} \to \mathbb{R}$ defined as
\begin{equation}\label{loss_c_def}
    \lossc(A, B) \coloneqq  \frac{1}{N^2}\left\|A-B\right\|_{\mathrm{F}}^2 = \frac{1}{N^2} \sum_{i, j} \left(A_{i,j} - B_{i, j}\right)^2,
\end{equation}
which measures how much $A$ is different from $B$. The disparity function is the building block of our loss function $\ell_{\mathrm{C}}$, already introduced in~\eqref{eq:classical-loss-1}. We remark that the pair $(\mysymmsetztwo{N}, \sqrt{\lossc(\cdot, \cdot)})$ determines a \textit{metric space} \cite{roman1992advanced} whose metric is induced
by the Frobenius norm.

We will implement the quantum loss as a quantum circuit, and for it, we define the observable $\mathcal{O}=\rvert 0\rangle^{\otimes (2k+1)}\langle0\rvert^{\otimes (2k+1)}$.
We then define the quantum disparity function as 
\begin{equation}
    \label{loss_q_def}
    \lossqgi(A,B) \coloneqq 1-\sqrt{\myprobz\left(
        \widehat{B} \cdot \widehat{A} \right)}\,.
\end{equation}

Using the definitions and facts above, we now prove that the classical and quantum disparity function are equivalent and we can use either one or the other to build a pertinent optimization problem. 

\begin{theorem}\label{lemma:classical=quantum}
Consider the adjacency matrices $A, B \in \mathcal{S}_N(\mathbb{Z}_2)$, and the disparity functions defined in~\eqref{loss_c_def} and \eqref{loss_q_def}. Then $\lossc(A, B) = \lossqgi(A,B)$. 
\end{theorem}

\begin{proof}
We expand \eqref{loss_q_def} to obtain
\begin{subequations}
\label{loss_graisom_expansion}
\begin{align}
    \lossqgi(A,B) &\underbrace{=}_{\text{Eq.~\eqref{probz_def}}}
    1-\left|\langle0\rvert^{\otimes (2k+1)} \widehat{B}\cdot \widehat{A}\, \rvert 0\rangle^{\otimes (2k+1)}\right|\\
    &\underbrace{=}_{\text{Prop.~\ref{hataplusb}, Lemma~\ref{lemma:exph}}} 1-
        \left|\bra{+}^{\otimes (2k+1)}
        \mathrm{cexp}(h(A +_2 B))
        \ket{+}^{\otimes (2k+1)}\right|\\
    &\underbrace{=}_{\text{Lemma~\ref{lemma:ctrl_U_inner_prod_plus}}} 1-
        \left|\frac{1}{2} + \frac{1}{2}\bra{+}^{\otimes 2k}
        \mathrm{exp}(h(A +_2 B))
        \ket{+}^{\otimes 2k}\right|\\
    &\underbrace{=}_{\text{Eq.~\eqref{simple_exph_exansion}}} 1 - \left|\frac{1}{2} + \frac{1}{2}\bra{+}^{\otimes 2k}
        \left(\sum_{i, j} (-1)^{A_{i, j} + B_{i, j}} \myouter{i, j}{i, j} \right)
        \ket{+}^{\otimes 2k}\right| \\ &= 1 - \Big|\underbrace{\frac{1}{2}
        + \frac{1}{2^{2k+1}} \sum_{i, j} (-1)^{A_{i, j} + B_{i, j}}}_{a}
        \Big| = 1-|a|. 
\end{align}
\end{subequations}
Now, one can check that $a\geq 0$, since $\sum_{i, j} (-1)^{A_{i, j} + B_{i, j}} \geq -2^{2k}$, and therefore we can take $a$ out of the absolute value, giving $1-|a| = 1-a$, yielding,
\begin{subequations}
\begin{align}
\lossqgi(A,B) & = \frac{1}{2} - \frac{1}{2^{2k+1}} \sum_{i, j} (-1)^{A_{i, j} + B_{i, j}} \underbrace{=}_{\text{Lemma~\ref{lemma:minus_one_pow_x_plus_y_equiv}}}
    \frac{1}{2} - \frac{1}{2^{2k+1}} \sum_{i, j} \left( 1 - 2(A_{i, j}-B_{i, j})^2\right)\\
    &=\frac{1}{N^2} \sum_{i, j} \left(A_{i, j}-B_{i, j}\right)^2
    =\lossc(A, B),
\end{align}   
\end{subequations}     
for all $A, B\in \mysymmsetztwo{N}$.
Hence $\lossqgi=\lossc$, consequently the pair $\left(\mysymmsetztwo{N}, \sqrt{\lossqgi(\cdot, \cdot)}\right)$ is a metric space.
\end{proof}

We can now build a quantum loss by considering
$\widehat{P_{\bar{p}}AP_{\bar{p}}^{\top}} = \check{P}_{\bar{p}}\cdot \widehat{A}\cdot \check{P}_{\bar{p}}^{\top}$ instead of $\widehat{A}$, obtaining the loss
\begin{subequations}
\label{loss_graisom_expansion_P}
\begin{align}
    \ell_{\mathrm{Q}}^{\mathrm{GI}}(P_{\bar{p}}; A,B) &=
    1 - \sqrt{\myprobz\left(\widehat{B}\cdot \widehat{P_{\bar{p}}A P_{\bar{p}}^{\top}}\right)}\underbrace{=}_{\text{Eq.~\eqref{probz_def}}}
    1-\left|\bra{0}^{\otimes (2k+1)} \widehat{B}\cdot \widehat{P_{\bar{p}}A P_{\bar{p}}^{\top}} \ket{0}^{\otimes (2k+1)}\right|\\
    &\underbrace{=}_{\text{Eq.~\eqref{hat_for_p}}} 1-\left|\bra{0}^{\otimes (2k+1)}
    \widehat{B}
    \cdot \left(\check{P} \widehat{A} \check{P}_{\bar{p}}^{\top}\right)
    \ket{0}^{\otimes (2k+1)} \right|
\end{align}
\end{subequations}
but by Lemma \ref{lemma-p_plus},
\begin{align}
    \check{P} \ket{0}^{\otimes (2k+1)} = \ket{0} \otimes \left(H^{\otimes 2k}\cdot P_{{\bar{p}}}^{\otimes 2} \cdot \ket{+}^{\otimes 2k}\right)= \ket{0} \otimes \ket{0}^{\otimes 2k}
\end{align}
which also applies to $\check{P}^{\top}$, so 
\begin{align}
    \label{loss_graisom_expansion_P:before_hat_expansion}
    \ell_{\mathrm{Q}}^{\mathrm{GI}}(P_{\bar{p}}; A,B) &= 1-\left|\bra{0}^{\otimes (2k+1)}
    \widehat{B} \cdot \check{P} \cdot \widehat{A}
    \ket{0}^{\otimes (2k+1)} \right| = \ell_{\mathrm{C}}(P_{\bar{p}}; A,B),
\end{align}
where the latter follows directly from Proposition~\ref{lemma:classical=quantum} substituting $A$ with $P_{\bar{p}} A P_{\bar{p}}^\top$.

Now, by expanding the $\widehat{(\cdot)}$ and the $\check{(\cdot)}$ in \eqref{loss_graisom_expansion_P:before_hat_expansion}, we obtain
\begin{equation}
    \ell_{\mathrm{Q}}^{\mathrm{GI}}(P_{\bar{p}}; A,B) = 1-\sqrt{\myprobz\left(U_{\mathrm{GI}}\right)}
\end{equation}
with the unitary operator $U_{\mathrm{GI}}$ defined as
\begin{equation}
    \label{loss_graisom_expansion_P:circ_unitary_op}
    U_{\mathrm{GI}} = 
    H^{\otimes (2k+1)} \cdot \mathrm{cexp}\left(h(B)\right)
    \cdot \left(\myiden{2} \otimes P_{{\bar{p}}}^{\otimes 2}\right)
    \cdot \mathrm{cexp}\left(h(A)\right) \cdot H^{\otimes (2k+1)}\,.
\end{equation}

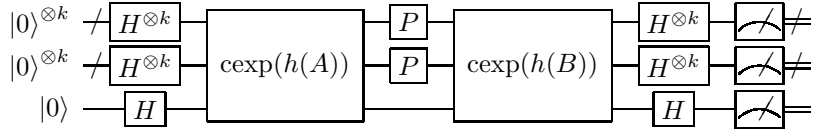
\begin{figure}
    \centering
    $\Qcircuit @C=0.5em @R=0.2em @!R {
            \nghost{ {q}_{1} :  } & \lstick{\ket{0}^{\otimes k}} & \qw {/} & \gate{H^{\otimes k}}
            & \qw & \multigate{2}{\mathrm{cexp}(h(A))}
            & \qw & \gate{P}
            & \qw & \multigate{2}{\mathrm{cexp}(h(B))}
            & \qw & \gate{H^{\otimes k}}
            & \qw & \meter & \cw{/} & \cw\\
            \nghost{ {q}_{2} :  } & \lstick{\ket{0}^{\otimes k}} & \qw{/} & \gate{H^{\otimes k}}
            & \qw & \ghost{\mathrm{cexp}(h(A))}
            & \qw & \gate{P}
            & \qw & \ghost{\mathrm{cexp}(h(B))}
            & \qw & \gate{H^{\otimes k}}
            & \qw & \meter & \cw{/} & \cw\\
            \nghost{ {q}_{3} :  } & \lstick{\ket{0}} & \qw & \gate{H}
            & \qw & \ghost{\mathrm{cexp}(h(A))}
            & \qw & \qw
            & \qw & \ghost{\mathrm{cexp}(h(B))}
            & \qw & \gate{H}
            & \qw & \meter & \cw & \cw
    }$
    \caption{The block circuit for the utility $\mathcal{L}_{\mathrm{Q}}^{\mathrm{GI}}$ given a fixed permutation $P$.} \label{circuit.qi}
\end{figure}

Considering now the utility
$\mathcal{L}_{\mathrm{Q}}^{\mathrm{GI}}(P_{\bar{p}}; A,B) = 1 - \ell_{\mathrm{Q}}^{\mathrm{GI}}(P_{\bar{p}}; A,B)$, then the graph isomorphism
problem can be formulated as 
\begin{equation}\label{problem:giq}
   \boxed{
    \max_{P_{\bar{p}} \in \Pi_N }\, \mathcal{L}_{\mathrm{Q}}^{\mathrm{GI}}(P_{\bar{p}}; A,B) \coloneqq
    \sqrt{\myprobz\left(U_{\mathrm{GI}}\right)},
    }
\end{equation}

whose utility can be evaluated in a quantum computer by the quantum circuit represented by the unitary matrix $U_{\mathrm{GI}}$. Problem~\eqref{problem:giq} is equivalent to Problem~\eqref{eq:sgi-classical}, in the sense that they have the same solutions in terms of permutation matrices $P_{\bar{p}}$, but the former can now be evaluated on a quantum computer (see Figure~\ref{circuit.qi} for the block circuit).

In the next section, we will show how to tune $P_{\bar{p}}$ via a quantum variational procedure, but now we focus on the subgraph isomorphism problem. 

\begin{remark}
We note that the simplification obtained in \eqref{loss_graisom_expansion_P:before_hat_expansion}
through Lemma~\ref{lemma-p_plus} assumes that $P$ is a permutation matrix, then so is $\myiden{2}\otimes P^{\otimes 2}$.
Later we are going to substitute $P$ with an Ansatz so we will return on this point to check the necessary conditions for this assumption to continue to hold. \hfill $\Box$
\end{remark}

\subsubsection{Sub-graph isomorphism case}
We extend the formulation of the loss to the case of the subgraph isomorphism, that is by allowing the adjacency matrix $B$ to have
arbitrary size $2^{k^\prime}\times 2^{k^\prime}$ with $k^\prime$ integer constrained to
$0\le k^{\prime}\le k$. The adjacency matrix $A$ instead is $N\times N=2^k\times 2^k$, so since $k^{\prime} \le k$, then 
the adjacency matrix $B$ is that of the subgraph that is matched against the set of subgraphs of the graph $\mathcal{A}$.

Consider the following operator
\begin{equation}
    Q=H^{\otimes(k-k^{\prime})} \otimes \myiden{2}^{\otimes k^\prime},
\end{equation}
which corresponds to applying the identity operator to the first $k^{\prime}$ qubits (from the least significative)
and the Hadamard operator to the subsequent $k-k^{\prime}$ qubits. 

Extend then the adjacency matrix $B$ by adding $N_{A}-N_{B}$ rows and columns of zeros, so to match the dimension of $A$, and therefore let
\begin{equation}
\Be = \left(\begin{array}{cc} B & \mathbb{O}_{N_B, N_A-N_B} \\ \mathbb{O}_{N_A-N_B, N_B} & \mathbb{O}_{N_A-N_B, N_A-N_B} \end{array}\right).
\end{equation}
Or equivalently, $\Be$ can be defined w.r.t. the following inner product
\begin{align}
    \bra{i}\Be\ket{j} = \begin{cases}
        B_{i, j} & \text{for $i, j \in \myfintsetp{k^\prime}$},\\
        0 & \text{otherwise.}
    \end{cases}
\end{align}

We define a new quantum disparity function as
\begin{equation}\label{loss_q_sgi_a}
    \lossqsgi(A,B) \coloneqq
    1 - \sqrt{\myprobz\left(
    \left(\myiden{2}\otimes Q^{\otimes 2}\right)
    \cdot \widehat{\Be}\cdot \widehat{A}
    \cdot \left(\myiden{2}\otimes Q^{\otimes 2}\right)
    \right)}\,.
\end{equation}
We expand Equation \eqref{loss_q_sgi_a} in Appendix \ref{appendix_subg_details}. The result reveals the relation between the quantum and the classical losses,
which is summarized in the following proposition.
\begin{theorem}\label{lemma:classical=quantum-2}
Consider the adjacency matrices $A \in \mathcal{S}_{N_A}(\mathbb{Z}_2)$ and $B \in\mathcal{S}_{N_B}(\mathbb{Z}_2)$ with $N_A \geq N_B$, and the disparity functions defined in~\eqref{loss_c_def} and \eqref{loss_q_sgi_a}.
Consider the block selection matrix $S = \begin{pmatrix}\mathbb{I}_{N_B} & \vert & \mathbb{O}_{N_B, N_A-N_B}\end{pmatrix}$. Then $\lossc(S A S^\top, B) = \lossqsgi(A,B)$. 
\end{theorem}

Now we proceed as in the previous section with the introduction of the permutation on $A$. So we obtain the following quantum loss,
\begin{align}
    \ell_{\mathrm{Q}}^{\mathrm{SGI}}(P_{\bar{p}}; A,B)
    &= 1-\left|\bra{0}^{\otimes (2k+1)} U_{\mathrm{SGI}} \ket{0}^{\otimes (2k+1)}\right|,
\end{align}
with the unitary operator $U_{\mathrm{SGI}}$ defined as
\begin{equation}
    U_{\mathrm{SGI}} = 
     \left(\myiden{2}\otimes Q^{\otimes 2}\right)
    \cdot \widehat{\Be}
    \cdot \left(\check{P}_{\bar{p}} \widehat{A} \check{P}_{\bar{p}}^{\top}\right)
    \cdot \left(\myiden{2}\otimes Q^{\otimes 2}\right).
\end{equation}
The details of the latter are expanded in Appendix \ref{appendix_subg_details}.

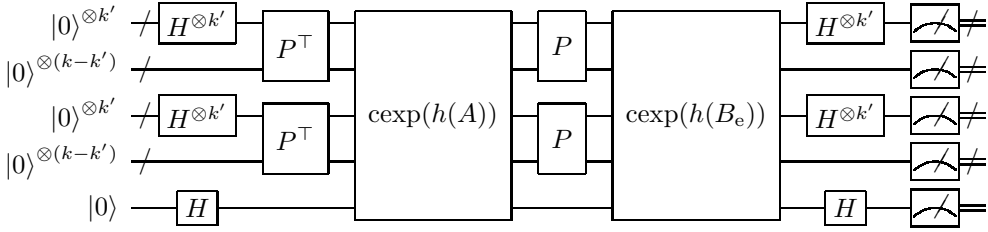
\begin{figure}
        \centering
    $\Qcircuit @C=0.5em @R=0.2em @!R {
            \nghost{ {q}_{1} :  } & \lstick{\ket{0}^{\otimes k^\prime}} & \qw {/} & \gate{H^{\otimes k^\prime}}
            & \qw & \multigate{1}{P^\top}
            & \qw & \multigate{4}{\mathrm{cexp}(h(A))}
            & \qw & \multigate{1}{P}
            & \qw & \multigate{4}{\mathrm{cexp}(h(\Be))}
            & \qw & \gate{H^{\otimes k^\prime}}
            & \qw & \meter & \cw{/} & \cw\\
            \nghost{ {q}_{2} :  } & \lstick{\ket{0}^{\otimes (k - k^\prime)}} & \qw{/} & \qw
            & \qw & \ghost{P^\top}
            & \qw & \ghost{\mathrm{cexp}(h(A))}
            & \qw & \ghost{P}
            & \qw & \ghost{\mathrm{cexp}(h(\Be))}
            & \qw & \qw
            & \qw & \meter & \cw{/} & \cw\\
            \nghost{ {q}_{3} :  } & \lstick{\ket{0}^{\otimes k^\prime}} & \qw{/} & \gate{H^{\otimes k^\prime}}
            & \qw & \multigate{1}{P^\top}
            & \qw & \ghost{\mathrm{cexp}(h(A))}
            & \qw & \multigate{1}{P}
            & \qw & \ghost{\mathrm{cexp}(h(\Be))}
            & \qw & \gate{H^{\otimes k^\prime}}
            & \qw & \meter & \cw{/} & \cw\\
            \nghost{ {q}_{2} :  } & \lstick{\ket{0}^{\otimes (k - k^\prime)}} & \qw{/} & \qw
            & \qw & \ghost{P^\top}
            & \qw & \ghost{\mathrm{cexp}(h(A))}
            & \qw & \ghost{P}
            & \qw & \ghost{\mathrm{cexp}(h(\Be))}
            & \qw & \qw
            & \qw & \meter & \cw{/} & \cw\\
            \nghost{ {q}_{4} :  } & \lstick{\ket{0}} & \qw & \gate{{H}}
            & \qw & \qw
            & \qw & \ghost{\mathrm{cexp}(h(A))}
            & \qw & \qw
            & \qw & \ghost{\mathrm{cexp}(h(\Be))}
            & \qw & \gate{H}
            & \qw & \meter & \cw & \cw
    }$
    \caption{The block circuit for the utility $\mathcal{L}_{\mathrm{Q}}^{\mathrm{SGI}}$ given a fixed permutation $P$.}\label{circuit.sqi}
\end{figure}

Considering now the utility
$\mathcal{L}_{\mathrm{Q}}^{\mathrm{SGI}}(P_{\bar{p}}; A,B) = 1 - \ell_{\mathrm{Q}}^{\mathrm{SGI}}(P_{\bar{p}}; A,B)$, then the subgraph isomorphism problem can be formulated as 
\begin{equation}\label{problem:sgiq}
   \boxed{
    \max_{P_{\bar{p}} \in \Pi_N }\, \mathcal{L}_{\mathrm{Q}}^{\mathrm{SGI}}(P_{\bar{p}}; A,B) \coloneqq
    \sqrt{\myprobz\left(U_{\mathrm{SGI}}\right)},
    }
\end{equation}
whose utility can be evaluated in a quantum computer by the quantum circuit represented by the unitary matrix $U_{\mathrm{SGI}}$. Problem~\eqref{problem:sgiq} is equivalent to Problem~\eqref{eq:sgi-classical}, in the sense that they have the same solutions in terms of permutation matrices $P_{\bar{p}}$, but the former can now be evaluated on a quantum computer (see circuit in Figure~\ref{circuit.sqi}).

\section{Solving the Sub-graph Isomorphism Problem}\label{sec:problem}

\subsection{Need for an Ansatz}

Having formulated the sub-graph isomorphism problem as~\eqref{problem:sgiq}, we proceed now on to solve it. As we notice, the decision variables are the permutation matrices $P_{\bar{p}}$ which need to be determined. While one can naively encode these matrices as SWAP gates (i.e., swapping two nodes at the time and reconstructing any permutation by a sufficiently large number of swaps), the resulting optimization problem would become combinatorial and very hard to solve in practice. The resulting problem would be in fact determining where to place SWAP gates and how many to use. 

Instead, here we look at an approximate Ansatz for the permutation matrices $P_{\bar{p}}$, say $\tilde{P}(\theta)$. Such structure is parametrized over continuous rotation angles $\theta$'s.
The approximation here is due to the fact that we may not look at all possible permutations, by truncating the number of iterations at a reasonable number, which seems to be working well in our numerical simulations.

\subsection{An approximate Ansatz}
\label{subsection:approx_ansatz}

We proceed constructively.
In this section, unless specified, we assume all permutations to be acting on $k$ qubits, so their underlying matrix is
an $N\times N$ permutation matrix with respect to the computational basis, where $N=2^k$.
As said, the idea here is that of substituting the permutation $P$ in the previous results with an Ansatz $\tilde{P}(\mathbf{\theta})$,
that is a parametric circuit generating permutations conditioned to the vector of continuous real variables $\mathbf{\theta}$.

Before that, we have to note that almost all results just obtained rely on the fact that $P$ is a permutation matrix.
For example, Lemma \ref{lemma-p_plus} applied to $\tilde{P}$ gives $\tilde{P}(\mathbf{\theta})\ket{+}^{\otimes k}=\ket{+}^{\otimes k}$
conditioned to $\tilde{P}$ being a permutation matrix.
This means that an arbitrary Ansatz would produce unexpected results due to the violation of the aforementioned condition.

However, we can observe that we can extend most of the previous required properties under linearity, for example in the case of Lemma \ref{lemma-p_plus} we see that
\begin{align}
    \label{ppluslinext}
    P_i \ket{+}^{\otimes k} = \ket{+}^{\otimes k} \implies& \left(\sum_{i}\left(\beta_i P_i\right)\right) \ket{+}^{\otimes k} = \sum_i\beta_i \ket{+}^{\otimes k},
\end{align}
where $\beta_i \in \mathbb{C}$ are arbitrary scalars and $P_i$ arbitrary $N\times N$ permutation matrices.
Also by requiring the unitarity of $\sum_{i}\left(\beta_i P_i\right)$ it follows that the RHS of \eqref{ppluslinext} is non-zero.
In essence, one can consider a linear combination of permutation matrices, instead of a single one, without losing the properties that one needs. 

Consider again the group-homomorphism $\varphi: \mathsf{S}_N \to \Pi_N$ defined in \eqref{p-P}.
We expand and formalize the concept by claiming that the preferred general form for the Ansatz, in the case of permutations of degree $N$, is the following,
\begin{align}
    \label{ansatz:general_form_i}
    \tilde{P}_{\mathrm{G}}(\mathbf{\theta}) =& \sum_{i=1}^{m} \alpha_i(\mathbf{\theta}) \cdot \varphi(p_{f(i)}),
\end{align}
where the functions $\alpha_i:\mathbb{R}^n \to \mathbb{C}$ map the Ansatz parameters to the coefficients of the permutations, and $\{p_{f(i)}\}_{i \in \myfintset{m} }$ is the set of vertex permutation functions, chosen by $f: \myfintset{m} \to \myfintset{N!}$ among the $N!$ possibilities. As for  $\tilde{P}_{\mathrm{G}}$ we require that 
\begin{enumerate}
    \item [C.1] 
    The expression $\tilde{P}_G(\mathbf{\theta})$ is a unitary operator.
    \item [C.2] Index
    $m \leq N!$ is some positive integer and $f: \myfintset{m} \to \myfintset{N!}$ is an injective mapping to the indices of the elements of $\symmg{N}$.
    \item [C.3]
    For each $i \in \myfintset{m}$, there exists a vector $\bar{\theta} \in \mathbb{R}^n$ such that $\left|\alpha_j(\bar{\theta})\right|\ne 0$ if and only if $i=j$,
    hence $\tilde{P}_{\mathrm{G}}(\bar{\theta})=\alpha_i(\bar{\theta}) \varphi(p_{f(i)}), \, |\alpha_i(\bar{\theta})|=1$, which means that $\tilde{P}_{\mathrm{G}}(\bar{\theta})$ is a permutation up to a global phase.
\end{enumerate}
Informally, condition C.2 (for $m< N!$) means that the Ansatz is not required to be able to generate all permutations for
the corresponding symmetric group.

On the basis of the conditions above, we prove that the linear superposition of permutations $\tilde{P}_{\mathrm{G}}$ is compatible with the loss minimization. For simplicity we consider only the graph isomorphism case.
\begin{theorem}
    \label{ansatz:prop_sol_preservation}
    Let $A, B \in \mysymmsetztwo{N}$ be the adjacency matrices of any pair of isomorphic graphs.
    Let $\lossqgi(\,\cdot\,; A, B)$ a graph isomorphism loss, with a solution $W$, that is $\lossqgi(W; A, B)=0$.
    Let $\tilde{P}_G$ a linear superposition of permutations satisfying the properties in its definition, such that $W$ is in the set of permutations generated by $\tilde{P}_G$.
    Then, there exists a $\theta \in \mathbb{R}^n$ such that $\tilde{P}_{\mathrm{G}}(\theta) = W$, possibly up to a global phase,
    and $\lossqgi\left(\tilde{P}_{\mathrm{G}}(\theta); A, B\right)=0$.
\end{theorem}
\begin{proof}
    Since $W$ is generated by $\tilde{P}_{\mathrm{G}}$, then, by the definition of $\tilde{P}_{\mathrm{G}}$,
    we know that there exists a $\theta \in \mathbb{R}^n$ such that $\tilde{P}_{\mathrm{G}}(\theta)=\alpha_i(\theta) W$,
    for some $i \in \myfintset{m}$. Also by the unitarity of $\tilde{P}_{\mathrm{G}}$, it follows that $|\alpha_i(\theta)|=1$.
    Hence,
    \begin{subequations}
    \begin{align}
        \lossqgi(\tilde{P}_{\mathrm{G}}(\theta); A,B) =& 1-\left|\bra{0}^{\otimes (2k+1)}
        \widehat{B} \cdot \left(\alpha_i(\theta)^2 \check{W}\right) \cdot \widehat{A}
        \ket{0}^{\otimes (2k+1)} \right|\\
        =& 1-\left|\alpha_i(\theta)\right|^2\cdot\left|\bra{0}^{\otimes (2k+1)}
        \widehat{B} \cdot \check{W} \cdot \widehat{A}
        \ket{0}^{\otimes (2k+1)} \right|= \lossqgi(Q; A, W)=0.
    \end{align}
    \end{subequations}
\end{proof}

We now design an instance of $\tilde{P}_{\mathrm{G}}$, consisting of the composition of an exponential number of permutations (not necessarily unique)
with respect to the number of elements of the vector $\mathbf{\theta}$, which is easier to handle that the general form, and can be easily generated on hardware. 
We start by setting $\mathbf{\theta}\in \mathbb{R}^n$ and $m=2^n$, with $n$ a free parameter of our choice.
Then, we consider a set of permutation functions $\left\{p_i\right\}_{i\in {n}} \subseteq \symmg{N}$, from the $N!$ possible
ones\footnote{Here, we will use the shorthand notation $p_i^k$, for any integer $k$, to indicate the $k$-fold
composition of $p_i$, i.e., $p_i^k = \underbrace{p_i \circ p_i \circ \ldots  \circ p_i}_{k \textrm{ times}}$.
And with this notation, $\varphi(p_i^k) = (\varphi(p_i))^k = P_i^k$}, such that they are self-inverse, i.e., $p_i \circ p_i = p_i^2 = \mathsf{e}_N$,
where $\mathsf{e}_N$ is the identity function (i.e., $\mathsf{e}_N \in \symmg{N}$, such that $\mathsf{e}_N\circ p=p\circ \mathsf{e}_N=p$ for all $p\in\symmg{N}$, mapping each vertex to itself, and $\varphi(\mathsf{e}_N) = \mathbb{I}_N$).
This latter condition is not restrictive, since any permutation can be written as composition of swaps, which fulfill
it~\footnote{A classical result in Group Theory says that given a positive integer $n$, the symmetric group $\symmg{n}$
is generated by the adjacent transpositions (here referred as swaps) \cite{Sagan2001}. These are $(1,2),(2, 3),\ldots,(n-1, n)$.},
but it is required for our technical development, as it will be clear shortly.  

We define then our preferred instance of the Ansatz as
\begin{equation}
    \label{ansatz:convenient_form}
    \tilde{P}(\mathbf{\theta}) = e^{\imath \phi} \prod_{i=1}^n \mathrm{exp}\left(-\imath\frac{\theta_i}{2}\varphi(p_i)\right).
\end{equation}
Thanks to the self-inverse property (see technical details in Lemma \ref{lemma_exp_perm}), we can expand the Ansatz as
\begin{subequations}
\begin{align}
\label{ansatz:convenient_form_as_cos_isin}
 \tilde{P}(\mathbf{\theta})    =& e^{\imath \phi} \prod_{i=1}^n \mathrm{exp}\left(-\imath\frac{\theta_i}{2}\varphi(p_i)\right) = e^{\imath \phi} \prod_{i=1}^n \left(\cos(\frac{\theta_i}{2})\myiden{N} -\imath\sin(\frac{\theta_i}{2})\varphi(p_i)\right)\\
    \label{ansatz:convenient_form_as_rot}
    =& e^{\imath \phi} \prod_{i=1}^n R_{P_i}(\theta_i),
\end{align}
\end{subequations}
where $R_{P_i}(\theta_i)$ is the generalized rotation gate with permutation $P_i$ and parameter $\theta_i$. 
A few remarks are now in order.

\begin{remark}
It is not difficult to see that $\tilde{P}$ has the form of the general $\tilde{P}_{\mathrm{G}}$ defined in Eq.~\eqref{ansatz:general_form_i}, it only suffices to expand Eq.~\eqref{ansatz:convenient_form_as_cos_isin} as
\begin{equation}
e^{\imath \phi} \prod_{i=1}^n \left(\cos(\frac{\theta_i}{2})\myiden{N} -\imath\sin(\frac{\theta_i}{2})\varphi(p_i)\right) = \sum_{j=1}^m \alpha_j(\theta) \varphi(p_j^{\circ}),
\end{equation}
where $p_j^{\circ}$ are compositions of multiples $p_i$'s so to match the left-hand side, and $m = 2^n$ as required. In addition, (C.1) $\tilde{P}(\theta)$ is a unitary operator\footnote{
The self-inverse condition on $p_i$ implies that the argument of the $\mathrm{exp}$ is a skew-hermitian matrix (i.e., $P_i^2=\myiden{N} \implies P_i=P_i^* \implies (\imath P_i)^*=-(\imath P_i)$), hence
$\mathrm{exp}\left(-\imath\frac{\theta_i}{2}\varphi(p_i)\right)$ is unitary. 
}; (C.2) holds by construction; and (C.3) for $\theta_i = \pi$, $\theta_j = 0$, for all $j \neq i$, then $\tilde{P}(\theta)$ is a permutation up to a global phase. \hfill $\Box$
\end{remark}

\begin{remark}
The permutations $p_i, p_j$ should be chosen such that they do not commute for $i\ne j$, so that for $P_i=\varphi(p_i), P_j=\varphi(p_j)$,
in general it does not follow that $\mathrm{exp}\left(-\imath\frac{\theta_i}{2}P_i\right)\cdot \mathrm{exp}\left(-\imath\frac{\theta_j}{2}P_j\right)
=\mathrm{exp}\left(-\frac{\imath}{2}(\theta_i P_i + \theta_j P_j)\right)$, and therefore the ``power or expression'' of the basis $p_i, p_j$ is ``larger'' than otherwise. For the same reason, the order of the factors $\mathrm{exp}$ above does matter. \hfill $\Box$
\end{remark}

\begin{remark}
    \label{ansatz:remark_intmul_pi}
    We illustrate an interesting fact following from Proposition \ref{ansatz:prop_sol_preservation} and the self-inverse property of $p_i$'s.
    Assume $\mathbf{\theta}=\pi \left(k_1, k_2, \ldots, k_n\right)^\top$ with $k_i \in \mathbb{Z}$, so that each $\theta_i$ is an integer multiple of $\pi$.
    Note that $\left|\cos\left(t\frac{\pi}{2}\right)\right|$ equals 1 for $t$ even and 0 otherwise, similarly
    $\left|\sin\left(t\frac{\pi}{2}\right)\right|$ is 1 when $t$ is odd and 0 otherwise.
    Also note that for any integer $t$, it follows that $P_i^{2t}=\left(P_i^2\right)^t=\myiden{N}$ and $P_i^{2t+1}=\left(P_i^2\right)^t\cdot P_i=P_i$.
    First we contend that for any integer $t$, $(-\imath)^t P_i^k=R_{P_i}(t\pi)$, indeed
    \begin{subequations}
    \begin{align}
        (-\imath)^t P_i^t =& e^{-\imath \frac{\pi}{2}t} P_i^t
        = \cos\left(\frac{\pi}{2}t\right)\cdot P_i^t-\imath \sin\left(\frac{\pi}{2}t\right)\cdot P_i^t\\
        =& \cos\left(\frac{\pi}{2}t\right)\cdot \myiden{N}-\imath \sin\left(\frac{\pi}{2}t\right)\cdot P_i
        = R_{P_i}(t\pi)\,,
    \end{align}
    \end{subequations}
where the last equation follows from the fact that when $t$ is even, only the first term is non-zero, and when $t$ is odd, only the second term is non-zero.
    So by extending the result to the factors of $\tilde{P}$ we obtain
    \begin{subequations}
    \begin{align}
        \tilde{P}(\theta) =& e^{\imath \phi} \prod_{i=1}^n R_{P_i}(\pi k_i)
        = e^{\imath \phi} \prod_{i=1}^n (-\imath)^{k_i} P_i^{k_i}
        = e^{\imath \phi^\prime} \prod_{i=1}^n P_i^{k_i}\,.
    \end{align}
    \end{subequations}
    The latter result means that when the parameters $\mathbf{\theta}$ are integer multiples of $\pi$, then $\tilde{P}(\mathbf{\theta})$ is a single permutation up to a
    global phase. This result is key for the optimization iteration since rounding the parameters to integer multiples of $\pi$ guarantees that
    the Ansatz implements a single permutation. Moreover, the latter allows a tractable classical determination of the permutation encoded by the
    values of the parameters. \hfill $\Box$
\end{remark}    
    
\begin{remark}    
    \label{ansatz:remark_pi_over_two_superpos}
    Let $\mathbf{\theta}=\frac{\pi}{2}(1,\,1,\ldots\,,1)$. By induction and considering the binary decomposition $j=\sum_{i=0}^{n-1} 2^{i}j_i$,
    with $j_i \in \{0, 1\}$, we obtain
	\begin{equation}
		\tilde{P}(\theta)
		= e^{\imath \phi} \prod_{i=1}^n \frac{1}{\sqrt{2}}\left(\myiden{N}-\imath P_i\right)
        = e^{\imath \phi^\prime} \frac{1}{\sqrt{2^{n}}}\sum_{j=0}^{2^n - 1} \prod_{i=1}^n P_i^{j_{i-1}},
	\end{equation}
    that is, for the specified $\mathbf{\theta}$, we obtain the perfect superposition of all permutations generable through the Ansatz. \hfill $\Box$
\end{remark}

The next step is that of obtaining the factors that form the Ansatz in \eqref{ansatz:convenient_form}, in terms of one- and two-qubit gates, that can be implemented on a near-term quantum computer. For the sake of simplifying notation, we introduce
an equivalence relation in the set of quantum circuits acting on $k$ qubits. We use again the concept of `doubling' introduced in Section~\ref{subsection:phy_distinguishability}.
So, given operators $U_a$ and $U_b$ acting on $k$ qubits, we define $U_a \sim U_b$ if and only if $\mathcal{D}(U_a)=\mathcal{D}(U_b)$. One can see that this definition satisfies the reflexivity, symmetric and transitivity properties.

To build the Ansatz~\eqref{ansatz:convenient_form}, we use now one and two-qubit gate primitives. First, we define the one-qubit gate primitive.

Consider the operator,
\begin{subequations}
\begin{align}
    H \cdot U_1(\lambda) \cdot H
    =& H \left(\myouter{0}{0} + e^{\imath\lambda}\myouter{1}{1}\right) H\\
    \sim& H \left(e^{-\imath\lambda/2} \myouter{0}{0} + e^{\imath\lambda/2}\myouter{1}{1}\right) H\\
    =& H \left(\cos(\lambda/2)\myiden{2} -\imath \sin(\lambda/2) \sigma_z\right) H = \cos(\lambda/2)\myiden{2} -\imath \sin(\lambda/2) \sigma_x = R_{\sigma_x}(\lambda),
\end{align}
\end{subequations}
so for nonnegative integers $k_1, k_2$ such that $k_1+k_2+1=k$,
\begin{subequations}
\begin{align}
    \myiden{2}^{\otimes k_1} \otimes \left(H \cdot U_1(\theta_i) \cdot H\right) \otimes \myiden{2}^{\otimes k_2}
    \sim& \cos(\lambda/2)\myiden{2}^{\otimes k} -\imath \sin(\lambda/2) \myiden{2}^{\otimes k_1} \otimes \sigma_x \otimes \myiden{2}^{\otimes k_2}\\
    \label{ansatz:gen_non_ent}
    =& R_{\myiden{2}^{\otimes k_1} \otimes \sigma_x \otimes \myiden{2}^{\otimes k_2}}(\lambda),
\end{align}
\end{subequations}
since $\myiden{2}^{\otimes k_1} \otimes \sigma_x \otimes \myiden{2}^{\otimes k_2}$ is a self-inverse permutation matrix, then
the latter has the form of the factors in \eqref{ansatz:convenient_form}, we call this the \textit{non-entangling primitive}.

We now construct an \textit{entangling primitive}. First we expand the controlled-phase gate
\begin{subequations}
\begin{align}
    \mathsf{C}(U_1(\lambda)) =& \myouter{0}{0}\otimes\myiden{2} + \myouter{1}{1}\otimes U_1(\lambda)\\
    \sim& e^{-\imath\lambda/2}\myouter{0}{0}\otimes\myiden{2}
    + \myouter{1}{1}\otimes\left(e^{-\imath\lambda/2} \myouter{0}{0} + e^{\imath\lambda/2}\myouter{1}{1}\right)\\
    =& \cos(\lambda/2) \myiden{2}^{\otimes 2} -\imath\sin(\lambda/2)\mathsf{C}(\sigma_z),
\end{align}
\end{subequations}
then, for hyper-parameter $a \in \{0, 1\}$, we introduce the operators,
\begin{subequations}
\begin{align}
    \mathrm{Ad}(H^a \otimes H^{1-a})\,\mathsf{C}(U_1(\lambda))
    \sim&\cos(\lambda/2) \myiden{2}^{\otimes 2} -\imath\sin(\lambda/2)
    \left(\mathrm{SWAP}^a \cdot \mathsf{C}(\sigma_x) \cdot \mathrm{SWAP}^a\right)\\
    \label{ansatz:gen_ent}
    =& R_{\mathrm{SWAP}^a \cdot \mathsf{C}(\sigma_x) \cdot \mathrm{SWAP}^a}(\lambda),
\end{align}
\end{subequations}
which again have the form of the factors in \eqref{ansatz:convenient_form}.
The gate $\mathrm{SWAP}^a$ is the classically controlled-swap gate, that is, given the classical control $a\in \{0, 1\}$,
the operator $\mathrm{SWAP}^a$ is the swap gate\footnotemark when $a=1$ and the identity $\myiden{2}^{\otimes 2}$ otherwise.
\footnotetext{
        $\vcenter{\Qcircuit @C=1.0em @R=0.2em @!R {
            \nghost{ {q}_{0} :  } & \lstick{ {q}_{0} :  } & \multigate{1}{\mathrm{SWAP}} & \qw\\
            \nghost{ {q}_{1} :  } & \lstick{ {q}_{1} :  } & \ghost{\mathrm{SWAP}} & \qw
        }}\,=\,
        \vcenter{\Qcircuit @C=1.0em @R=0.2em @!R {
            \nghost{1} & \lstick{ {q}_{0} :  } & \targ & \ctrl{1} & \targ & \qw\\
            \nghost{1} & \lstick{ {q}_{1} :  } & \ctrl{-1} & \targ & \ctrl{-1} & \qw
        }}$
}
The latter construction is justified by the fact that the multiset of eigenvalues of the $\mathrm{CNOT}$ operator is
$\{1^{(3)}, -1^{(1)}\}$, then $\mathsf{C}(\sigma_z)$ corresponds to the diagonal matrix of such eigenvalues, so 
the expression $H^a \otimes H^{1-a}$ constitutes the eigenvectors that determine the control and target.

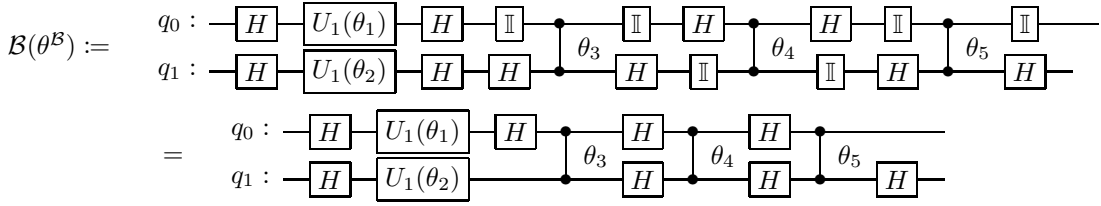
\begin{figure}
    \centering
    $\mathcal{B}(\mathbf{\theta}^\mathcal{B})\coloneqq\vcenter{
    \Qcircuit @C=1.0em @R=0.2em @!R {
        \nghost{ {q}_{1} :  } & \lstick{ {q}_{0} :  } & \gate{{H}} & \gate{{U_1}(\theta_1)} & \gate{{H}} & \gate{\mathbb{I}} & \ctrl{1}
        & \dstick{\theta_3} \qw & \gate{\mathbb{I}}& \gate{{H}} & \control \qw
        & \dstick{\theta_4} \qw & \gate{{H}} & \gate{\mathbb{I}} & \ctrl{1}
        & \dstick{\theta_5} \qw & \gate{\mathbb{I}} & \qw & \qw\\ 
        \nghost{ {q}_{2} :  } & \lstick{ {q}_{1} :  } &  \gate{{H}} & \gate{{U_1}(\theta_2)}
        & \gate{{H}} & \gate{{H}} & \control \qw
        & \qw & \gate{{H}} & \gate{\mathbb{I}} & \ctrl{-1}
        & \qw & \gate{\mathbb{I}} & \gate{{H}} & \control \qw
        & \qw & \gate{{H}} & \qw
    }}$ \\ \vspace*{0.2cm}
     $=\vcenter{
    \Qcircuit @C=1.0em @R=0.2em @!R {
        \nghost{ {q}_{1} :  } & \lstick{ {q}_{0} :  } & \gate{{H}} & \gate{{U_1}(\theta_1)}
        & \gate{{H}} & \ctrl{1}
        & \dstick{\theta_3} \qw & \gate{{H}} & \control \qw
        & \dstick{\theta_4} \qw & \gate{{H}} & \ctrl{1}
        & \dstick{\theta_5} \qw & \qw & \qw\\ 
        \nghost{ {q}_{2} :  } & \lstick{ {q}_{1} :  } & \gate{{H}} & \gate{{U_1}(\theta_2)}
        & \qw & \control \qw
        & \qw & \gate{{H}} & \ctrl{-1}
        & \qw & \gate{{H}} & \control \qw
        & \qw & \gate{{H}} & \qw
    }}$
    \caption[An example of a basic block.]{An example of a basic block $\mathcal{B}(\mathbf{\theta})$ for the proposed Ansatz composed of two non-entangling primitives,
    and three entangling ones. The parameter $\theta^{\mathcal{B}} = (\theta_1, \ldots, \theta_5)$ collects all the rotations, and the
    $\vcenter{\Qcircuit @C=1.0em @R=1.5em{& \ctrl{1} & \qw\\ & \control \qw & \ustick{\theta_i} \qw}}\,$
    represents the $\mathsf{C}(U_1(\lambda))$ operator.}
    \label{ansatz:block}
\end{figure}

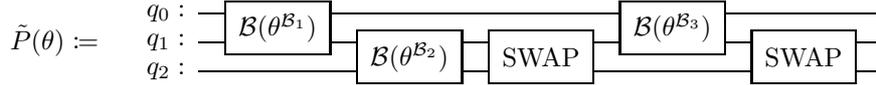
\begin{figure}
    \centering
    $\tilde{P}(\mathbf{\theta})\coloneqq
    \vcenter{
    \Qcircuit @C=1.0em @R=0.2em @!R {
        \nghost{ {q}_{1} :  } & \lstick{ {q}_{0} :}
        & \multigate{1}{\mathcal{B}(\theta^{\mathcal{B}_1})}
        & \qw & \qw
        & \multigate{1}{\mathcal{B}(\theta^{\mathcal{B}_3})}
        & \qw & \qw\\
        \nghost{ {q}_{2} :  } & \lstick{ {q}_{1} :}
        & \ghost{\mathcal{B}(\theta^{\mathcal{B}_1})}
        & \multigate{1}{\mathcal{B}(\theta^{\mathcal{B}_2})}
        & \multigate{1}{\mathrm{SWAP}}
        & \ghost{\mathcal{B}(\theta^{\mathcal{B}_3})}
        & \multigate{1}{\mathrm{SWAP}} & \qw\\
        \nghost{ {q}_{3} :  } & \lstick{ {q}_{2} :}
        & \qw & \ghost{\mathcal{B}(\theta^{\mathcal{B}_2})}
        & \ghost{\mathrm{SWAP}}
        & \qw & \ghost{\mathrm{SWAP}} & \qw
    }}$
     \caption{A composition of basic blocks forming the Ansatz $\tilde{P}(\mathbf{\theta})$ for a three-qubit case.
     We call this topology \textit{circular}.
     The parameter $\theta = (\theta^{\mathcal{B}_1}, \theta^{\mathcal{B}_2}, \theta^{\mathcal{B}_3})$ collects all the rotation parameters of the circuit. The construction
    $(\mathrm{SWAP}\otimes \myiden{2})\cdot(\mathcal{B}(\theta^{\mathcal{B}_3}) \otimes \myiden{2})\cdot(\mathrm{SWAP}\otimes \myiden{2})$
    corresponds to having the block $\mathcal{B}(\theta^{\mathcal{B}_3})$ acting on the first and last qubits.}
    \label{ansatz:first_3q}
\end{figure}

Finally, we generate the proposed Ansatz using the primitives in \eqref{ansatz:gen_non_ent} and \eqref{ansatz:gen_ent}. One can define an arbitrary block composing non-entangling and entangling primitives (as in Figure~\ref{ansatz:block}) acting on a two-qubit circuit, and then use this block as a two-qubit gate to use to generate the Ansatz (in Figure~\ref{ansatz:first_3q} we propose a possible construction for the three-qubit case). 

We note here that the primitives are the only fixed elements of the construction, while the topology is free to be chosen. Different topologies will generate different permutations.

\begin{remark}
    We note that for the case of the graph isomorphism, the implementation of a generic permutation Ansatz may not be necessary.
    This because we can exclude from the solutions the permutations that associate vertices of different degree, where the degree of a vertex is the
    number of edges incident on it. A possible solution consists of partitioning the vertices by degree and rearranging the adjacency matrices such that
    vertices belonging to the same class (degree) are contiguous. If we assume that there are $\mathsf{P}$ partitions, each having a power-of-two cardinality, then the
    Ansatz can be built as the tensor product of Ansatze $\bigotimes_{i=1}^\mathsf{P} \tilde{P}_i(\mathbf{\theta}_i)$.
    In the general case, where one or more partitions do not present a power-of-two cardinality, the strategy introduced before involving ancillary singleton vertices
    can be employed in this case as well.
    The reduction in the number of \textit{conjugancy classes} the Ansatz is required to implement is then substantial, since the number of conjugancy classes for a symmetric group
    of degree $n$ is given by the partition function $\mathsf{p}(n)$ \cite{andrews1994number}, which has the asymptotic form
    $\mathsf{p}(n) \sim \frac{1}{4n\sqrt{3}}\mathrm{exp}(\pi\sqrt{\frac{2n}{3}})$ (Ramanujan).
\end{remark}

\begin{remark}
\rev{We cite here also the work~\cite{SeelbachBenkner2021} where a parametric cyclic expansion is proposed to tackle optimization over permutations in a quantum algorithm. As in our case, the expansions uses a basis of $2$-cycles, i.e., SWAP gates, to encode arbitrary permutations. In general, the approach remains different from ours, but it has the same spirit of encoding a suitable Ansatz instead of introducing constraints. }
\end{remark}

\subsection{The proposed algorithm}

We propose an algorithm for the sub-graph isomorphism problem based on the framework for representing, permuting and comparing adjacency matrices,
developed in the first part of the present work. The procedure is illustrated in pseudo-code in Algorithm \ref{variational_algo}. The graph isomorphism
problem is a special case of the latter so we are going to concentrate on the former.

The input consists of the adjacency matrices corresponding to the source and pattern graph, respectively. These are assumed having size $N\times N$ and $N_B\times N_B$,
respectively, where $N=2^k$ and $N_B=2^{k^\prime}$, with $k>k^\prime$. We further encode the topological structure of the Ansatz into the hyper-parameters $\mathfrak{G}$.
As explained in Section \ref{subsection:approx_ansatz}, the Ansatz is not expected to cover the entire search space of permutations,
so re-iterating the procedure with different sets of hyper-parameters (typically random) has the benefits of extending the coverage of the generated permutations.

The algorithm consists of two main phases, namely, the preparatory and the iterative steps.

In the {\bf preparatory step}, first we prepare the block selection matrix $S$, which is used, as explained in Section \ref{subsection:classical_loss},
to obtain the partial permutation for the classical evaluation of the discrepancy.
We also prepare the utility function $\mathcal{L}_{\mathrm{Q}}^{\mathrm{SGI}}(\mathbf{\theta})$, which, in practice consists of a circuit generator depending
on $\mathbf{\theta}$ and the chosen Ansatz for the permutation superposition and its topology $\mathfrak{G}$.
The initial $\mathbf{\theta}$ is drawn from a uniform random distribution on the subset $\Theta :=\left\{\omega\in \mathbb{R}^n\middle|\omega_i \in \left[0, \pi\right]\right\}$, whose justification is a consequence of Remark \ref{ansatz:remark_intmul_pi}.

The {\bf iterative phase} essentially alternates between an iteration of the stochastic gradient descent (SGD) algorithm (or a variant of the latter),
acting on the parameters $\theta$ of the Ansatz, and a classical sampler that evaluates a number of solutions sampled from the superposition represented by the current parameters.
The choice of SGD is supported by the recent theoretical work on SGD for hybrid quantum-classical optimization \cite{sgdhybrudqoptim}.
In relation to the latter, our formulation presents the sequence of random variables
$\left\{\widetilde{\nabla}^{(t_1)} \mathcal{L}_{\mathrm{Q}}^{\mathrm{SGI}}(\mathbf{\theta})\right\}$,
these are the estimators of the gradient. The convergence is guaranteed when such estimators are \textit{unbiased}, that is
$\mathbb{E}\left[ \widetilde{\nabla}^{(t_1)} \mathcal{L}_{\mathrm{Q}}^{\mathrm{SGI}}(\mathbf{\theta})\right]=\nabla \mathcal{L}_{\mathrm{Q}}^{\mathrm{SGI}}(\mathbf{\theta})$.
In \cite{sgdhybrudqoptim} it is shown that the estimation of the expectation via a finite number of measurements, constitutes an unbiased estimator, hence the SGD is justified.
In our implementation SGD makes use of a numerical gradient, but we do not exclude the possibly of extending the method to the gradient evaluated with the parameter shift
rule \cite{crooks2019gradients}.

Since a run of the algorithm is expected to obtain at most a solution, the overall procedure must be interpreted as a sampler of solutions rather than an exact solver.
Also, the effect of the random initialization of the parameters $\mathbf{\theta}$ and the noise of the device, is that of making the algorithm hit a potentially different solution
at each execution.

The algorithm then needs a mechanism to round the obtained parameters $\theta$ to integers of $\pi$ to ``project'' them onto the permutation space. We use a probabilistic rounding technique, variations of which are widely used in convex relaxations and signal processing, see e.g.,~\cite{Luo2010}. Loosely speaking, each continuous $\theta_i$ can be used to determine the probability of which of the two surrounding integers of $\pi$ is selected (see Appendix~\ref{app-probabilistic} for the formal details).  The probabilistic procedure draws (classically) a number of possible integer solutions that are used to obtain permutations, and
which are tested against the input adjacency matrices. The algorithm stops when we reach the maximum number of steps or at least a solution is found.

The algorithm described in Algorithm \ref{variational_algo} can be reiterated a number of times according to some user-defined rule (changing the initial $\theta$ or the topology $\mathfrak{G}$, or the choice of Ansatz), each iteration can potentially produce a solution of the sub-graph isomorphism problem.

\newcommand{\ptildetheta}{\widetilde{P}_{\mathfrak{G}}(\mathbf{\theta})}
\begin{algorithm}
\footnotesize
    \caption{The pseudocode for a single iteration of the algorithm}
    \label{variational_algo}
    \SetAlgoLined
    \KwData{Adjacency matrices $A$ and $B$ for graphs $\mathcal{A}$ and $\mathcal{B}$, respectively.
    Permutation Ansatz hyper-parameters $\mathfrak{G}$.}
    \tcp{The matrices $A$ and $B$ have sizes $N\times N$ and $N_B\times N_B$,$N=2^k$, $N_B=2^{k^\prime}$, $k^\prime < k$}
    \SetKwData{Maxsteps}{maxsteps}\SetKwData{Samples}{samples}
    \SetKw{Break}{break}
    \KwResult{$R$\tcp*{Set of partial permutation matrices}}
    $R \leftarrow \emptyset$\;
    $S \leftarrow \begin{pmatrix}\mathbb{I}_{N_B} & \vert & \mathbb{O}_{N_B, N_A-N_B}\end{pmatrix}$
    \tcp*{Prepare the block selection matrix (section \ref{subsection:classical_loss})}
    \tcp{Prepare the utility $\mathcal{L}_{\mathrm{Q}}^{\mathrm{SGI}}$ circuit}
    \smallskip
    $\mathcal{L}_{\mathrm{Q}}^{\mathrm{SGI}}(\mathbf{\theta}) \longleftarrow
    \vcenter
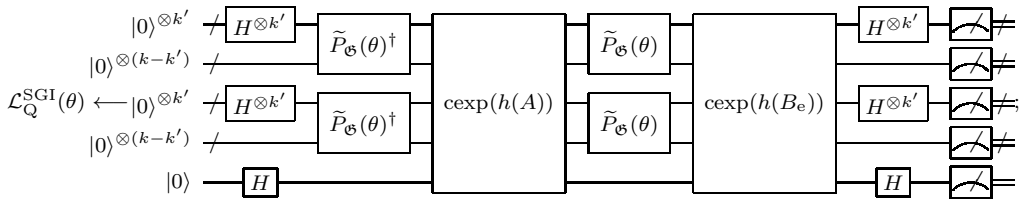
{
    \Qcircuit @C=0.5em @R=0.2em @!R {
            \nghost{ {q}_{1} :  } & \lstick{\ket{0}^{\otimes k^\prime}} & \qw {/} & \gate{H^{\otimes k^\prime}}
            & \qw & \multigate{1}{\ptildetheta^\dagger}
            & \qw & \multigate{4}{\mathrm{cexp}(h(A))}
            & \qw & \multigate{1}{\ptildetheta}
            & \qw & \multigate{4}{\mathrm{cexp}(h(\Be))}
            & \qw & \gate{H^{\otimes k^\prime}}
            & \qw & \meter & \cw{/} & \cw\\
            \nghost{ {q}_{2} :  } & \lstick{\ket{0}^{\otimes (k - k^\prime)}} & \qw{/} & \qw
            & \qw & \ghost{\ptildetheta^\dagger}
            & \qw & \ghost{\mathrm{cexp}(h(A))}
            & \qw & \ghost{\ptildetheta}
            & \qw & \ghost{\mathrm{cexp}(h(\Be))}
            & \qw & \qw
            & \qw & \meter & \cw{/} & \cw\\
            \nghost{ {q}_{3} :  } & \lstick{\ket{0}^{\otimes k^\prime}} & \qw{/} & \gate{H^{\otimes k^\prime}}
            & \qw & \multigate{1}{\ptildetheta^\dagger}
            & \qw & \ghost{\mathrm{cexp}(h(A))}
            & \qw & \multigate{1}{\ptildetheta}
            & \qw & \ghost{\mathrm{cexp}(h(\Be))}
            & \qw & \gate{H^{\otimes k^\prime}}
            & \qw & \meter & \cw{/} & \cw\\
            \nghost{ {q}_{2} :  } & \lstick{\ket{0}^{\otimes (k - k^\prime)}} & \qw{/} & \qw
            & \qw & \ghost{\ptildetheta^\dagger}
            & \qw & \ghost{\mathrm{cexp}(h(A))}
            & \qw & \ghost{\ptildetheta}
            & \qw & \ghost{\mathrm{cexp}(h(\Be))}
            & \qw & \qw
            & \qw & \meter & \cw{/} & \cw\\
            \nghost{ {q}_{4} :  } & \lstick{\ket{0}} & \qw & \gate{{H}}
            & \qw & \qw
            & \qw & \ghost{\mathrm{cexp}(h(A))}
            & \qw & \qw
            & \qw & \ghost{\mathrm{cexp}(h(\Be))}
            & \qw & \gate{H}
            & \qw & \meter & \cw & \cw
    }}$\;
    \smallskip
    $\mathbf{\theta} \leftarrow \mathrm{SampleUniform}(\left[0, \pi\right]^n)$\tcp*{Sample initial $\mathbf{\theta} \in \mathbb{R}^n$ from $\left[0, \pi\right]^n$}
    \For{$t_1\leftarrow 1$ \KwTo \Maxsteps}{
        $\mathbf{\theta} \leftarrow \mathbf{\theta} -\eta \widetilde{\nabla}^{(t_1)} \left(-\mathcal{L}_{\mathrm{Q}}^{\mathrm{SGI}}(\mathbf{\theta})\right)$
        \tcp*{Update $\mathbf{\theta}$ with SGD step}
        \tcp{Obtain vector of distances of $\mathbf{\theta}_i/\pi$ from closest even integer}
        $\mathbf{p} \leftarrow \left(\Lambda\left(\theta_i/\pi\right)\right)_i$\;
        \tcp{Classical sampling loop}
        \For{$t_2\leftarrow 1$ \KwTo \Samples}{
            $\mathbf{u} \leftarrow \mathrm{SampleUniform([0, 1]^n)}$\;
            $\mathbf{g} \leftarrow \left(\chi_{[0,\infty)}(p_i - u_i) \right)_i$
            \tcp*{Sample vector in $\{0, 1\}^n$ according to probabilities $\mathbf{p}$}
            $P \leftarrow \widetilde{P}_{\mathbf{h}}(\pi \cdot \mathbf{g})$\tcp*{Obtain a classical permutation, note $\pi \cdot \mathbf{g} \in \{0, \pi\}^n$}
            $d \leftarrow \left\|SP A P^{\top}S^\top - B\right\|^2_F$\tcp*{Check discrepancy classically}
            \If{$d=0$}{ $R \leftarrow R \cup \{SP\}$\tcp*{New solution} }
        }
        \If{$|R|\ne 0$}{\Break\tcp*{If solutions set not empty interrupt}}
    }
\end{algorithm}

\begin{remark}
    \label{tildep_tp_tildep_remark}
    We note that in Algorithm \ref{variational_algo} the Ansatz is applied as $\tilde{P}(\mathbf{\theta}) \otimes \tilde{P}(\mathbf{\theta})$.
    The theoretical construction of the approach represented in Figure \ref{circuit.qi} assumes that $P$ is a single permutation.
    Now, $\tilde{P}(\mathbf{\theta})$ is in general a linear combination of permutations $P_i$ so if for simplicity we assume the set of permutations
    $\left\{\myiden{N}, P_1\right\}$ then $\tilde{P}(\mathbf{\theta})=\alpha_0(\mathbf{\theta}) \myiden{N} + \alpha_1(\mathbf{\theta}) P_1$.
    But
    \begin{subequations}
    \begin{align}
        \tilde{P}(\mathbf{\theta}) \otimes \tilde{P}(\mathbf{\theta})
        =& \alpha_0(\mathbf{\theta})^2 \cdot \myiden{N} \otimes \myiden{N}
        + \alpha_0(\mathbf{\theta}) \alpha_1(\mathbf{\theta}) \cdot \myiden{N} \otimes P_1\\
        &+ \alpha_0(\mathbf{\theta}) \alpha_1(\mathbf{\theta}) \cdot P_1\otimes \myiden{N}
        + \alpha_1(\mathbf{\theta})^2 \cdot P_1 \otimes P_1\,,
    \end{align}
    \end{subequations}
    which is not exactly what we were expecting since the terms $\myiden{N} \otimes P_1$ and $P_1\otimes \myiden{N}$ correspond classically to apply
    different permutations to the rows and columns of the adjacency matrix.
    The implementation of an Ansatz that produces only superpositions of permutations of the form $P_i \otimes P_i$ is possible but it requires additional entanglement.
    However, the unwanted configurations peak at $\mathbf{\theta}=\frac{\pi}{2}\left(1,\,1,\,\ldots,\,1\right)$ and vanish when all elements of $\mathbf{\theta}$
    are integer multiples of $\pi$, as consequence of Proposition \ref{ansatz:prop_sol_preservation}, so we have not observed any issue for the implementation of the algorithm. 
    We reserve further investigations on the matter to future extensions of this research. \hfill $\Box$
\end{remark}

\section{Experiments}\label{sec:experiments}

\begin{figure}
    \centering
    \scalebox{0.4}{\input{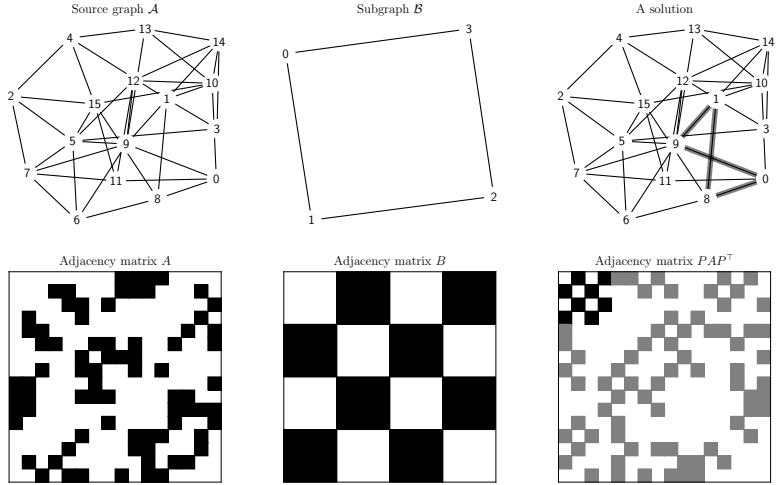}}
    \caption{The data used for testing the subgraph isomorphism algorithm (first two columns). In the last column, we highlight a solution
    obtained by means of the proposed method. Here the adjacency matrices are depicted as a plot in which black rectangles represent
    non-zero entries of the matrix. In the third column, note that the top-left sub-matrix (highlighted) of the permuted adjacency matrix
    $PAP^\top$ corresponds to the adjacency matrix of the pattern graph. Notably the loss depends exclusively on the highlighted sub-matrix.}
    \label{experiment:sgi_i}
\end{figure}

We pass now to the investigation of our algorithms in numerical examples. The experiments consist of subgraph and graph isomorphism problems featuring graphs of different
sizes and patterns. An example of the considered graphs is depicted in Figure~\ref{experiment:sgi_i}.

We generate the input graphs as Erd\"{o}s-R\'{e}nyi random graphs \cite{erdos59a}. 
The algorithm devised in Algorithm \ref{variational_algo} is implemented in Python using the Qiskit framework, and for the evaluation of the circuits we
use the Qiskit simulator\footnote{\rev{The implementation is available at \url{https://github.com/qiskit-community/subgraph-isomorphism}.}}. The classical optimizer is the stochastic gradient descent with momentum, configured with $\lambda=0.9$ (the momentum term).
The learning rate is set to $\eta=0.1$ and $\epsilon$ for the numerical gradient is set to $\epsilon=0.1$, for all experiments.
Moreover, the number of shots for estimating the expectation is set to 1024.

Each iteration of the algorithm is limited to a maximum of $128$ steps (gradient descent iterations), so we define a run as non-convergent when we have not obtained any
valid solution within that limit; the procedure also stops when we hit a solution. We remark that $128$ steps are a small number in general for gradient descent,
and used here to keep the total quantum simulation time on our classical computer manageable\footnote{For instance, the simulation for the experiment presenting the greatest
permutation space size took approximately 40 hours on an Intel(R) Core(TM) i7-8850H CPU.}.

\begin{figure}
    \centering
    \scalebox{0.4}{
    \input{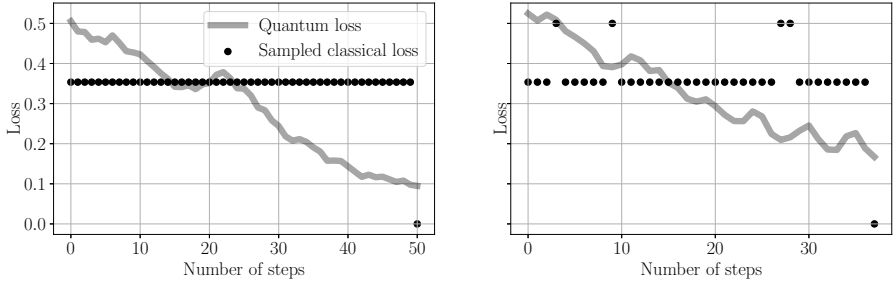}}
    \caption{The evolution of the loss for two runs of the SGI problem depicted in Figure \ref{experiment:sgi_i}.
    The points marked with `$\bullet$' represent the classical loss for the best sample obtained classically
    considering the distribution induced by the parameters of the Ansatz. Note that the classical loss \rev{is not continuous in the parameters} because it is related to a single solution (sampled) \rev{and may vary more erratically}, whereas the quantum loss can be interpreted as a weighted superposition of solutions \rev{and it is the one that is optimized for}.}
    \label{experiment:loss_plot}
\end{figure}

The evolution of the loss for two runs of the SGI problem depicted in Figure \ref{experiment:sgi_i} is presented in Figure~\ref{experiment:loss_plot}. The points marked with `$\bullet$' represent the classical loss for the best sample obtained classically considering the distribution induced by the parameters of the Ansatz. Note that the classical loss \rev{is not continuous in the parameters} because it is related to a single solution (sampled) \rev{and may vary more erratically}, whereas the quantum loss can be interpreted as a weighted superposition of solutions \rev{and it is the one that is optimized for}.

We then perform several numerical experiments. The tests are divided into two classes, namely, the search and convergence tests. Search tests are meant to evaluate the ability of the algorithm to identify solutions
when there is no guarantee that the Ansatz can represent the permutations related to the matches.
In the case of the convergence tests, we are just interested in understanding the ability of the algorithm to converge to a solution when there is the guarantee that the Ansatz
can represent such solution or one of its symmetries. To do so we choose random permutations from those implementable by the selected Ansatz to create the problems.

For the Ansatz we use the circular topology presented in Figure~\ref{ansatz:first_3q}. In the presentation of the algorithm it has been explained that to favor
a greater coverage of the search space, the hyper-parameters for the Ansatz (i.e., the descriptors of its topology) should be randomized at each run of the algorithm.
In relation to the latter we choose a slightly different approach, so for the search tests we keep the Ansatz topology constant (circular form) and we apply a
random pre-permutation to the source graph before each run. The pre-permutation is then composed with the solution to form the solution with respect to the original graph.

For each case, we run the algorithm $100$ times, also we sample $64$ solutions classically after each iteration of the SGD. Each execution starts with a random vector of
parameters $\mathbf{\theta}$.

The summary of the statistics for the experiments is presented in Table~\ref{experiment:table}. We number the different problems by $R$, and add a letter, e.g., $1a, 2a$, in the case of a different realization of the same instance.
         The total number of parameters $\theta$ for the Ansatz is indicated as $n$. For the space size, we apply the formula $\frac{N_A!}{(N_A-N_B)!}$,
         reported in Section~\ref{subsection:classical_loss}. When possible we obtain classically (brute force) the number of unique solutions and the
         total number of symmetries of the latter (third column).

As we can observe, with a small number of parameters $15-20$, we find solutions of the problems in many cases. The percentage of convergent runs at $128$ steps are smaller in the search tests (as one can expect, since the Ansatz is not necessarily in the permutation space), but still reasonable for such a small number of parameters. Notice that a non-convergent run, offering an approximate solution, may still be useful in practice when dealing with approximate sub-graph isomorphism and motif detection. 

Remark the differences within the realizations $1a, 1b, 1c$ and $2a, 2b, 2c, 2d$: even having the same number of nodes of the underlying graphs, the number of unique solutions and symmetries renders solving the problem easier (when this number is high) or harder (when this number is low).  

For the convergence test, the number of convergent runs is above $80\%$ in most cases, even with the small number of parameters, and for the largest problem instance, it is still a reasonable $52$-$65\%$ and basically limited by the $128$ steps cap.

\begin{table}
\caption{Table of the statistics for the experiments for different problems $R$. For each problem, we run the algorithm $100$ times.
         The total number of parameters $\theta$ for the Ansatz is indicated as $n$. For the space size, we apply the formula $\frac{N_A!}{(N_A-N_B)!}$,
         reported in Section~\ref{subsection:classical_loss}. When possible we obtain classically (brute force) the number of unique solutions and the
         total number of symmetries of the latter (third column).}
    \label{experiment:table}

    \begin{center}
        \scalebox{0.8}{
        \begin{tabular}{c c c c c c c c c c} 
        \toprule
        \multicolumn{10}{c}{\bf Search tests} \\
        \toprule
        \multicolumn{3}{c}{Problem} & Space size & Unique sol.s; symm.& Unique sol.s & N. of par.ters; & Convergent & Avg SGD & Max SGD \\ 
        $R$ & $N_A$ & $N_B$ & & (brute force) & found & qubits; depth & runs [\%] & steps & steps\\
        \toprule
        $1a$ & $8$ & $4$ & 1680 & 2; 8 & 2 & 15; 7; 505 & 42 & 9.2 & 52 \\ 
        $1b$ & $8$ & $4$ & 1680 & 6; 24 & 6 & 15; 7; 512 & 75 & 5.6 & 45 \\ 
        $1c$ & $8$ & $4$ & 1680 & 3; 12 & 3 & 15; 7; 510 & 49 & 5.2 & 31 \\ \midrule
        $2a$ & $16$ & $4$ & 43680 & 36; 288 & 32 & 20; 9; 1841 & 79 & 7.8 & 71 \\ 
        $2b$ & $16$ & $4$ & 43680 & 29; 116 & 22 & 20; 9; 1844 & 65 & 16.5 & 113 \\ 
        $2c$ & $16$ & $4$ & 43680 & 53; 212 & 50 & 20; 9; 1849 & 83 & 13.6 & 119 \\ 
        $2d$ & $16$ & $4$ & 43680 & 26; 104 & 20 & 20; 9; 1847 & 56 & 18.7 & 114 \\ 
        \toprule
        \end{tabular}}
        \\ \vspace{5mm}
        \scalebox{0.8}{
        \begin{tabular}{c c c c c c c c c c} 
        \toprule
        \multicolumn{10}{c}{\bf Convergence tests} \\
        \toprule
         \multicolumn{3}{c}{Problem} & Space size & Unique sol.s; symm.& Unique sol.s & N. of par.ters; & Convergent & Avg SGD & Max SGD \\ 
         $R$ & $N_A$ & $N_B$ & & (brute force) & found & qubits; depth & runs [\%] & steps & steps\\
         \toprule
         $3$ & $8$ & $4$ & $1680$ & 18; 36 & 2 & 15; 7; 506 & 100 & 2.2 & 16 \\ 
         $4$ & $16$ & $4$ & $43680$ & 109; 218 & 2 & 20; 9; 1847 & 100 & 5.6 & 41 \\ 
         $5a$ & $16$ & $8$ & $\approx 5.2\cdot 10^8$ & N.A. & 1 & 20; 9; 1998 & 65 & 52.6 & 127 \\ 
         $5b$ & $16$ & $8$ & $\approx 5.2\cdot 10^8$ & N.A. & 1 & 20; 9; 1884 & 52 & 66.0 & 109 \\ 
         \toprule
         $6a$ & $8$ & $8$ & 40320 & 1; 4 & 1 & 15; 7; 480 & 94 & 23.4 & 99 \\
         $6b$ & $8$ & $8$ & 40320 & 1; 2 & 1 & 15; 7; 474 & 88 & 35.4 & 123 \\
         $6c$ & $8$ & $8$ & 40320 & 1; 2 & 1 & 15; 7; 487 & 84 & 34.2 & 125 \\ 
         $6d$ & $8$ & $8$ & 40320 & 1; 4 & 1 & 15; 7; 481 & 84 & 30.7 & 124 \\
         \toprule
        \end{tabular}}
    \end{center}
   \end{table}

\subsection{A closer look}

\begin{figure}
    \centering
    \scalebox{0.425}{
    \input{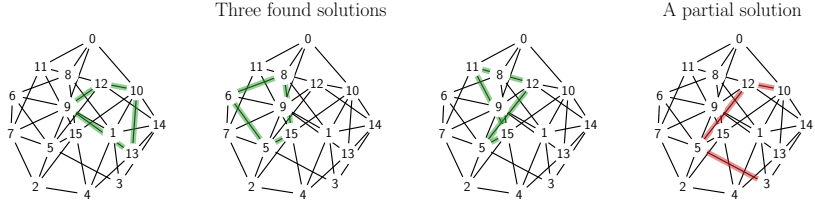}}
    \caption{\rev{Instances of the solutions obtained with our method for the setting $2a$. Out of the $32$ unique solutions found, we report three. We also report a case of non-convergence of the algorithm (a so-called partial solution). Since $2a$ is a search test, the reason for non-convergence is here due to the fact that the chosen ansatz does not cover the permutations that are required. }}
    \label{solutions:sgi}
\end{figure}

\begin{figure}
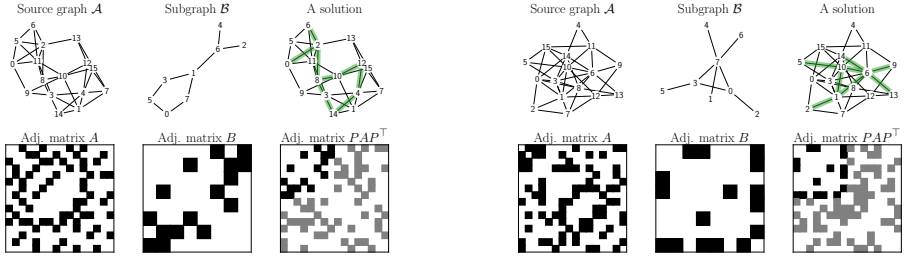

    \centering
    \scalebox{0.2}{
    \input{experiment_5a.pgf}}
    \scalebox{0.2}{
    \input{experiment_5b.pgf}}
    \caption{\rev{Instances of the solutions obtained with our method for the settings $5a$ and $5b$. }}
    \label{solutions:sgi-5}
\end{figure}

\begin{figure}
    \centering
    \scalebox{0.5}{
    \input{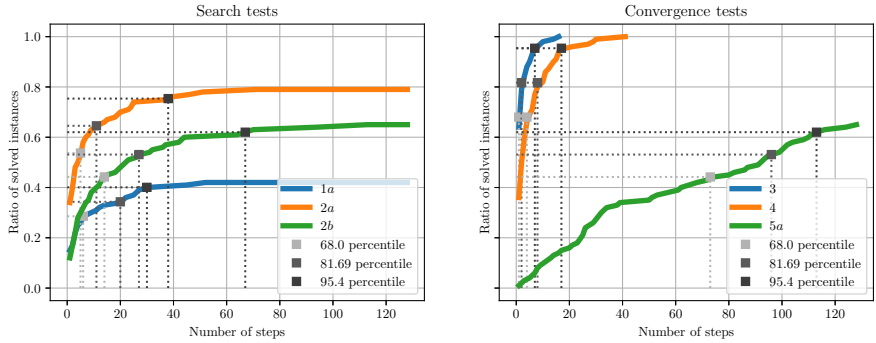}}
    \caption{\rev{How the number of SGD steps affects the ratio of solved instances? We report three cases for each test regime. In the search tests, we see that $128$ steps are abundant, and quickly the number of solved instances reaches its maximum, after which the ansatz is too small to cover the permutations required. In the convergent tests, we see that the blocking feature can be the number of steps, especially when graphs are bigger, but otherwise we can solve $100\%$ of instances very quickly. }}
    \label{comparison:sgi}
\end{figure}

\rev{In Table~\ref{experiment:table}, we also report the number of qubits needed to run the sub-graph isomorphism circuit, as well as the depth. The latter is measured using Qiskit's \texttt{QuantumCircuit.depth()} method on the transpiled circuit featuring the alphabet $\{U_3, CX\}$. Such process is stochastic, so the reported depth is an average.

As we see the depth grows faster than the number of qubits, which it has to be expected. The dependence of the number of qubits with the graph dimension is logarithmic, but not the dependence of the depth, which is quadratic in the total number of vertices for Qiskit.   

In Figure~\ref{solutions:sgi}, we report instances of the solutions obtained with our method for the setting $2a$. Out of the $32$ unique solutions found, we report three. We also report a case of non-convergence of the algorithm (a so-called partial solution). Since $2a$ is a search test, the reason for non-convergence is here due to the fact that the chosen ansatz does not cover the permutations that are required. However, this non-convergent case, leading to a partial solution, is still relevant in some applications, like approximate sub-graph isomorphism and motif detection. 

In Figure~\ref{solutions:sgi-5}, we report the solutions obtained for the settings $5a$ and $5b$, which are our largest settings. Here it is worth noticing that, despite the very large space size ($\approx 5.2 \cdot 10^8$), we are able to find solutions with only $20$ parameters and with a very limited number of SGD iterations ($\leq 128$). 

Finally, Figure~\ref{comparison:sgi} captures how the number of SGD steps affects the ratio of solved instances within a setting. We report three cases for each test regime. In the search tests, we see that $128$ steps are abundant, and quickly the number of solved instances reaches its maximum, after which the ansatz is too small to cover the permutations required. In the convergent tests, we see that the blocking feature can be the number of steps, especially when graphs are bigger, but otherwise we can solve $100\%$ of instances very quickly. These plots are instrumental to pick the best number of steps and possibly the ansatz. 
}

While we leave for future investigations the study of the influence of the parameters choice and topology, the presented results already depict a promising and sound performance of the proposed algorithm. 

\subsection{Classical and quantum comparisons}

\begin{figure}
    \centering
    \scalebox{0.45}{
    \input{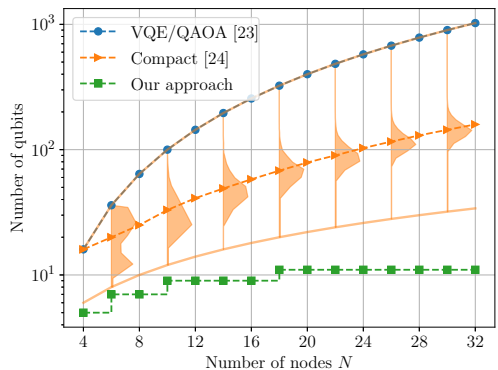}}
    \caption{\rev{Quantum approaches comparison for the graph isomorphism problem. We report the qubit requirements varying the number of nodes of the graph for the VQE/QAOA solution of~\cite{Chatterjee2021}, the compressed approach of~\cite{Kenneth2015}, and our approach. For~\cite{Kenneth2015}, we plot the min, max, and median and selected distributions.}}
    \label{qq-comparison:sgi}
\end{figure}

\rev{We close with some comparisons. 

We have used a classical implementation of the algorithm V2F~\cite{Cordella2004} provided in the Python NetworkX package to compare our results with a classical solver (other very performant solvers are for example nauty/Traces~\cite{McKay2014}). The classical algorithm finds all the expected solutions in fractions of seconds for the considered small instances. In this regime, classical algorithms are more competitive, yet -- as already mentioned in the introduction -- the maximum instance considered in the literature features graphs of $10^6$ nodes, while we could encode graphs with $10^{15}$ nodes on a $100$-qubit computer. This could give a clear advantage to our algorithm on very large graphs. 

As for quantum approaches, we have studied the graph isomorphism QUBO/QAOA solutions of~\cite{Chatterjee2021} and \cite{Kenneth2015} and analyzed their qubit requirements as well as iteration counts varying the number of nodes in the graph. In Figure~\ref{qq-comparison:sgi}, we plot the number of qubits that their Hamiltonian requires. For~\cite{Chatterjee2021}, the number is deterministic and exactly $N^2$, while for~\cite{Kenneth2015} it depends on how the graph is connected. Hence, for the latter, we plot minimal and maximal requirements (which can be computed analytically as $N+2$, and $N^2$) as well as the median over $2^{15}$ random graph realizations (which the authors empirically estimate to go as $\sim N \log_2 N$). We plot also the distribution of requirements for selected nodes. And finally, we compare it with our approach. 

As we can see, our approach is a staircase graph, since we need to add virtual nodes whenever $N$ is not a power of $2$, but it is by far the most efficient for qubit resources. In addition, considering the number of iterations that are required by VQE, the authors in~\cite{Chatterjee2021} report at least $600$ iterations or more to reach a solution for a $4$-node graph, and one can expect similar figures when implementing~\cite{Kenneth2015} on a gate-based quantum simulator. Since we have less than $128$ iterations across all the graphs that we have studied, our approach seems to be the most competitive.

}

\section{Conclusions}\label{sec:conclusions}
We have discussed a novel algorithm to solve sub-graph isomorphism problems on a gate-based quantum computer. The algorithm scales better than current literature with the number of nodes of the underlying graphs (namely, it requires a number of qubits that is a logarithmic function of the nodes), and it could provide a solid use case for quantum computing on hard optimization problems in the medium term. 

Further research avenues encompass generalizations of the problem (e.g., digraph isomorphism, common induced subgraph, labeled edges and nodes), improved compilation of the resulting circuit, enhanced classical solvers, and a comparative study of different Ansatze.

\newpage
\appendix
\section{Probabilistic rounding procedure}\label{app-probabilistic}

We discuss here the probabilistic rounding procedure for sampling integers multiple of $\pi$, starting from a continuous parameter $\theta$.
Specifically, we are interested exclusively in the distinction even/odd multiples.
We let $\theta \in \mathbb{R}$. 

We introduce the \textit{triangle wave function} $\Lambda: \mathbb{R}\to [0, 1]$ which we define as follows
\begin{align}
    \Lambda(x) \coloneqq& \left| \left(\lfloor x \rfloor\mod{2}\right) - \left(x\mod{1}\right) \right|,
\end{align}
where $(\cdot)\mod{1}$ is the \textit{division modulo 1}. 
For any $x\in \mathbb{R}$, consider the decomposition $x=z+r$, with $z\in\mathbb{Z}$ and
$r\in[0, 1)$, then we define
\begin{align}
    x \mod{1} \coloneqq& x - \lfloor x \rfloor\,,
\end{align}
so $z=\lfloor x \rfloor$ and $r=x\mod{1}$.
The triangle wave function, as defined above, can be interpreted as the distance of a point $x\in \mathbb{R}$ on the real line to the closest
even integer.
We prove the latter fact.
It can be shown that a real number $x\in \mathbb{R}$ can be decomposed as either $x=2k+r$ or $x=(2k+1)+r$,
with $k\in \mathbb{Z}$ and $r\in [0, 1)$.
Consider the case $x=2k+r$, then $\Lambda(2k+r)=\left| \left(2k \mod{2}\right) - r \right|=r$, that is the distance from $x$ to $2k$
(also note that any other even $2(k+t)$ with $t \in \mathbb{Z}\setminus \{0\}$ is further from $x$).
Similarly, for $x=(2k+1)+r$ we see that $\Lambda((2k+1)+r)=1-r$, that is the distance from $x$ to $2(k+1)$.

We define the \textit{indicator function} $\chi_X(x): Y \to \{0, 1\}$ for a subset $X \subseteq Y$ as
\begin{align}
    \chi_X(x) \coloneqq \begin{cases}
        1, & x\in X,\\
        0, & x\notin X\,.
    \end{cases}
\end{align}
Using the concepts just introduced, we will be considering a Bernoulli random variable with parameter $p=\Lambda(\theta/\pi) \in [0,1]$ and a sample $u$ from the uniform
distribution on $[0, 1]$. We note that the expression $\chi_{[0,\infty)}(p - u)$ corresponds to obtaining $1$ whenever $u \le p$ and 0 otherwise,
that is a sample from the Bernoulli random variable.

So the rounding procedure rounds $\theta$ to the closest odd integer multiple of $\pi$ (or equivalently to $\pi$) if $\chi_{[0,\infty)}(p - u)$ is $1$,
and to the closest even multiple of $\pi$ (or equivalently to $0$) otherwise.

\section{Details of the sub-graph isomorphism loss}
\label{appendix_subg_details}
We proceed by expanding the quantum disparity function in \eqref{loss_q_sgi_a} to reveal its relation with the classical disparity $\lossc$, so
\begin{align}
    \label{loss_subgraisom_intermediate_i}
    \lossqsgi(A,B)
    \underbrace{=}_{\text{Prop.~\ref{hataplusb}, Lemma~\ref{lemma:exph}}}& 1-
        \left|\bra{+} \otimes \bra{\psi}^{\otimes 2} \mathrm{cexp}(h(A +_2 \Be)) \ket{+} \otimes \ket{\psi}^{\otimes 2}\right|
\end{align}
where we have defined $\ket{\psi}$ as:
\begin{align}
    \ket{\psi}=& H^{\otimes k} Q \ket{0}^{\otimes k} = H^{\otimes k} \left(H^{\otimes(k-k^{\prime})} \otimes \myiden{2}^{\otimes k^\prime}\right) \ket{0}^{\otimes k} =  \left(\myiden{2}^{\otimes(k-k^{\prime})} \otimes H^{\otimes k^\prime}\right) \ket{0}^{\otimes k}=\ket{0}^{\otimes (k-k^{\prime})} \otimes \ket{+}^{\otimes k^{\prime}}\,.
\end{align}
Also, note that,
\begin{subequations}
\begin{align}
    \bra{\psi} \ket{i}
    =& \left(\bra{0}^{\otimes (k-k^{\prime})} \otimes \bra{+}^{\otimes k^{\prime}}\right) \ket{i_{k-1} i_{k-2} \ldots i_0}\\
    =& \bra{0}^{\otimes (k-k^{\prime})} \ket{i_{k-1} i_{k-2} \ldots i_{k^\prime}}
    \cdot \bra{+}^{\otimes k^{\prime}} \ket{i_{k^{\prime}-1} i_{k^{\prime}-2} \ldots i_0}= \begin{cases}
        \frac{1}{\sqrt{2^{k^\prime}}}, & \text{if $i_{k-1}=i_{k-2}=\cdots=i_{k^\prime}=0$,}\\
        0, & \text{otherwise},
    \end{cases}
\end{align}
\end{subequations}
similarly for $\bra{\psi} \ket{j}$, so
\begin{align}
    \label{subg_isom_loss:psi_ij}
    \bra{\psi}^{\otimes 2} \ket{i, j} = \myinnerp{\psi}{i}\cdot\myinnerp{\psi}{j}= \begin{cases}
        \frac{1}{2^{k^\prime}}, & \text{if $\sum_{t=k^\prime}^{k-1}(i_t + j_t)=0$,}\\
        0, & \text{otherwise},
    \end{cases}
\end{align}
additionally, $\bra{i, j}\left(\ket{\psi}^{\otimes 2}\right)=\bra{\psi}^{\otimes 2} \ket{i, j}$.
Continuing from \eqref{loss_subgraisom_intermediate_i} using the same approach applied in \eqref{loss_graisom_expansion}, we have
\begin{subequations}
\label{dummy}
\begin{align}
    \lossqsgi(A,B) &\underbrace{=}_{\text{Lemma~\ref{lemma:ctrl_U_inner_prod_plus}}} 1-
        \left|\frac{1}{2} + \frac{1}{2}\bra{\psi}^{\otimes 2}
        \mathrm{exp}(h(A +_2 \Be))
        \ket{\psi}^{\otimes 2}\right|\\
    &= 1 - \left|\frac{1}{2} + \frac{1}{2}\bra{\psi}^{\otimes 2k}
        \left(\sum_{i, j} (-1)^{A_{i, j} + B_{i, j}} \myouter{i, j}{i, j} \right)
        \ket{\psi}^{\otimes 2k}\right|\\
    &\underbrace{=}_{\text{Eq.~\eqref{subg_isom_loss:psi_ij}}} 1 - \left|\frac{1}{2}
        + \frac{1}{2^{2k^{\prime}+1}} \sum_{i=0, j=0}^{2^{k^\prime}} (-1)^{A_{i, j} + B_{i, j}}
        \right|\\
    &\underbrace{=}_{\text{Lemma~\ref{lemma:minus_one_pow_x_plus_y_equiv}}}
    \frac{1}{2^{2k^{\prime}}} \sum_{i=0, j=0}^{2^{k^\prime}} \left(A_{i, j}-B_{i, j}\right)^2
    =\lossc(S A S^\top, B),
\end{align}
\end{subequations}
Hence we proved the claim of Proposition \ref{lemma:classical=quantum-2}.

Now we proceed as in the GI case with the introduction of the permutation on $A$. So we obtain the following quantum loss,
\begin{subequations}
\begin{align}
    \label{subg_isom_loss:perm_expansion}
    \ell_{\mathrm{Q}}^{\mathrm{SGI}}(P_{\bar{p}}; A,B)
    &= 1-\left|\bra{0}^{\otimes (2k+1)}
    \left(\myiden{2}\otimes Q^{\otimes 2}\right)
    \cdot \widehat{\Be}
    \cdot \widehat{P_{\bar{p}}A P_{\bar{p}}^{\top}}
    \cdot \left(\myiden{2}\otimes Q^{\otimes 2}\right)
    \ket{0}^{\otimes (2k+1)}\right|\\
    \label{subg_isom_loss:perm_expansion_i}
    &\underbrace{=}_{\text{by Eq.~\eqref{hat_for_p}}} 1-\left|\bra{0}^{\otimes (2k+1)}
    \left(\myiden{2}\otimes Q^{\otimes 2}\right)
    \cdot \widehat{\Be}
    \cdot \left(\check{P}_{\bar{p}} \widehat{A} \check{P}_{\bar{p}}^{\top}\right)
    \cdot \left(\myiden{2}\otimes Q^{\otimes 2}\right)
    \ket{0}^{\otimes (2k+1)} \right| \\ &= 1-\left|\bra{0}^{\otimes (2k+1)} U_{\mathrm{SGI}} \ket{0}^{\otimes (2k+1)}\right|,
\end{align}
\end{subequations}
with the unitary operator $U_{\mathrm{SGI}}$ defined as
\begin{equation}
    U_{\mathrm{SGI}} = 
     \left(\myiden{2}\otimes Q^{\otimes 2}\right)
    \cdot \widehat{\Be}
    \cdot \left(\check{P}_{\bar{p}} \widehat{A} \check{P}_{\bar{p}}^{\top}\right)
    \cdot \left(\myiden{2}\otimes Q^{\otimes 2}\right).
\end{equation}

\section{Additional proofs}
\label{appendix_more_proofs}
\begin{lemma}
    \label{lemma_exp_perm}
    Let $\theta \in \mathbb{R}$ and $P$ be an $n\times n$ permutation matrix such that $P^2=\myiden{n}$, then
    \begin{align}
        \mathrm{exp}\left(\imath\theta P\right) = \cos(\theta)\myiden{n} +\imath\sin(\theta)P\,.
    \end{align}
\end{lemma}
\begin{proof}
    First we notice that the power-series of the exponential over the real numbers, fixed the argument $x\in \mathbb{R}$,
    is \textit{absolutely convergent} (by the ratio test). Then the exponential series over the set of all $n\times n$ matrices \textit{converges in norm} \cite{rossmann2006lie}.
    Also $P^2=\myiden{n}$ implies $P^{2t}=\myiden{n}$ and $P^{2t+1}=P$ for all $t\in \mathbb{Z}$.
    So,
    \begin{align}
        \mathrm{exp}\left(\imath\theta P\right) = 
        \sum_{t=0}^{\infty} \frac{\left(\imath\theta P\right)^t}{t!}
        = \myiden{n}\sum_{t=0}^{\infty} \frac{(-1)^t}{(2t)!} \theta^{2t}
        + \imath P \sum_{t=0}^{\infty} \frac{(-1)^t}{(2t+1)!} \theta^{2t+1}
        =\cos(\theta)\myiden{n} +\imath\sin(\theta)P\,.
    \end{align}
\end{proof}

\begin{proof}[Proof of Proposition \ref{dbl_hat_not_faithful_proof}]
    By Proposition \ref{hataplusb}, the definition of $\mathcal{D}$, and the interchange law of the tensor product we have
    \begin{subequations}
    \begin{align}
        \mathcal{D}\left(\widehat{A +_2 B}\right) =& \mathcal{D}\left(\widehat{A}\cdot \widehat{B}\right)
        = \overline{\widehat{A}\cdot \widehat{B}} \otimes \widehat{A}\cdot \widehat{B}\\ 
        =& \left(\overline{\widehat{A}} \otimes \widehat{A}\right) \cdot \left(\overline{\widehat{B}} \otimes \widehat{B}\right)=\mathcal{D}\left(\widehat{A}\right) \cdot \mathcal{D}\left(\widehat{B}\right)
        = \mathcal{D}\left(\widehat{B}\right) \cdot \mathcal{D}\left(\widehat{A}\right)\,,
    \end{align}
    \end{subequations}
    where the commutativity follows from \eqref{doubling_pres_abel}, finally,
    \begin{align}
        \mathcal{D}\left(\widehat{\mathbb{O}_N}\right) =& \mathcal{D}\left(\mathbb{I}_{2N^2}\right)
        = \mathbb{I}_{2N^2} \otimes \mathbb{I}_{2N^2}\,,
    \end{align}
    thus $\mathcal{D}\left(\widehat{(\cdot)}\right)$ is a homomorphism.

    To determine the kernel of the composition $\mathcal{D}\left(\widehat{(\cdot)}\right)$, we have to determine the subset of $A\in \mysymmsetztwo{N}$
    such that $\mathcal{D}\left(\,\widehat{A}\,\right)=\mathcal{D}(\mathbb{I}_{2N^2})$ (since the latter is the identity element in the codomain of $\mathcal{D}$).
    That is,
    \begin{align}
        \label{double_hat_kernel_i}
        \mathcal{D}\left(\,\widehat{A}\,\right)=\mathcal{D}(\myiden{2N^2})
        \iff& \mathrm{cexp}\left(\overline{h(A)}\right) \otimes \mathrm{cexp}\left(h(A)\right) = \mathcal{D}(\myiden{2N^2})
    \end{align}
    but by Lemma \ref{lemma:exph} it follows that
    \begin{subequations}
    \begin{align}
        \mathrm{cexp}\left(\overline{h(A)}\right) \otimes \mathrm{cexp}\left(h(A)\right) =&
        \mathrm{cexp}\left(h(A)\right)^{\otimes 2}\\
        =& \left(\myiden{N^2} \oplus \mathrm{exp}\left(h(A)\right)\right)^{\otimes 2}
    \end{align}
    \end{subequations}
    now, let $V=\mathrm{exp}\left(h(A)\right)$ and $I=\myiden{N^2}$, then by applying \eqref{otimes_oplus_relation} twice we obtain
    \begin{subequations}
    \begin{align}
        \left(\myiden{N^2} \oplus \mathrm{exp}\left(h(A)\right)\right)^{\otimes 2} =& \left(I \oplus V\right)^{\otimes 2}\\
        \cong& I^{\otimes 2} \oplus (V \otimes I) \oplus (I \otimes V) \oplus V^{\otimes 2}
    \end{align}
    \end{subequations}
    by also considering that $(V \otimes I) \cong (I \otimes V)$, implication \eqref{double_hat_kernel_i} becomes
    \begin{subequations}
    \begin{align}
        \mathcal{D}\left(\,\widehat{A}\,\right)=\mathcal{D}(\myiden{2N^2})
        \iff& 
        I^{\otimes 2} \oplus (V \otimes I) \oplus (I \otimes V) \oplus V^{\otimes 2} = \bigoplus_{i=1}^{4} I^{\otimes 2}\\
        \iff& I \otimes V  = I^{\otimes 2}\\
        \label{double_hat_kernel_last}
        \iff& V = I.
    \end{align}
    \end{subequations}
    Thus, \eqref{double_hat_kernel_last} implies $\mathrm{exp}\left(h(A)\right)=\myiden{N^2}$, that is
    \begin{align}
        \sum_{i, j} (-1)^{A_{i, j}} \myouter{i, j}{i, j}=\myiden{N^2} \iff& A_{i, j}=0\quad\forall\,i,j\,,
    \end{align}
    hence $A=\mathbb{O}_{N}$, so $\mathrm{ker}\, \mathcal{D}\left(\,\widehat{(\cdot)}\,\right) = \{\mathbb{O}_N\}$
    as claimed.
\end{proof}

\section*{Acknowledgement}
This study has received funding from the Disruptive Technologies Innovation Fund (DTIF), by Enterprise
Ireland, under project number DTIF2019-090 (project QCoIR), and also
supported by IBM Quantum and Mastercard Ireland.
We are highly thankful to Dr. Claudio Gambella, Dr. Martin Mevissen (IBM Research Europe - Dublin) and Prof. Jiri Vala (Maynooth University - Dep. of Theoretical Physics)
for their precious suggestions. 

\bibliographystyle{IEEETran}
\bibliography{refs}

\end{document}